\documentclass[12pt,a4paper]{article}

\usepackage{amsthm,amsmath,natbib,amssymb,amsfonts,bm}
\usepackage{graphicx}
\usepackage{placeins}

\def\hang{\hangindent\parindent}
\def\rf{\par\noindent\hang}

\newtheorem{theorem}{Theorem}

\newtheorem*{theorem*}{Theorem}

\theoremstyle{definition}

\theoremstyle{remark}

\DeclareMathAlphabet{\mathpzc}{OT1}{pzc}{m}{it}

\newcommand{\SSB}{\text{\small SSB}}
\newcommand{\SSW}{\text{\small SSW}}

\newcommand{\CP}{\text{\small CP}}
\newcommand{\CPK}{\text{\small CPK}}
\newcommand{\T}{\text{\small T}}

\def\hang{\hangindent\parindent}
\def\rf{\par\noindent\hang}

\begin{document}

\baselineskip=20pt

\begin{center}
%\noindent
{\bf \Large The impact of a Hausman pretest, applied to panel data, on the coverage probability of confidence intervals}
\end{center}

%\bigskip

%Running title: CONFIDENCE INTERVALS AND PRIOR INFORMATION

\bigskip

\begin{center}
{\bf \large PAUL KABAILA$^*$, RHEANNA MAINZER {\normalsize AND} DAVIDE FARCHIONE}
\end{center}

\medskip

\begin{center}
%{\large
{\sl Department of Mathematics and Statistics, La Trobe University, Australia}
%}
\end{center}

\vspace{1cm}

\noindent \textbf{Summary} \ \ In the analysis of panel data that includes a time-varying covariate, a Hausman pretest is commonly used to decide
whether subsequent inference is made using the random effects model or the fixed effects model. We consider the effect of this
pretest on the coverage probability of a confidence interval for the slope parameter.
We prove three new finite sample theorems that
 make it easy to assess, for a wide variety of circumstances, the effect of the
Hausman pretest on the minimum coverage probability of this confidence interval.
Our results show that for the small levels of significance of the Hausman pretest commonly used in applications,
the minimum coverage probability of the confidence interval for the slope parameter can
be far below nominal.

\bigskip

%\begin{center}

%\end{center}

\vspace{6.5cm}

\noindent * Corresponding author. Department of Mathematics and Statistics,
La Trobe University, Victoria 3086, Australia. Tel.: +61 3 9479 2594; fax +61 3 9479 2466.
{\sl E-mail address:} P.Kabaila@latrobe.edu.au.

\newpage

\begin{center}
\textbf{1. INTRODUCTION}
\end{center}

%\medskip

\noindent In the analysis of panel data that includes a time-varying covariate, a preliminary Hausman (1978) test is commonly used to decide
whether subsequent inference is made using the random effects model or the fixed effects model. If the Hausman pretest rejects the null hypothesis of no correlation between
the random effect and time-varying covariate then the fixed effects model is chosen for subsequent inference, otherwise the
random effects model is chosen. This preliminary model selection procedure has been widely used in econometrics (see e.g. Wooldridge, 2002 and Baltagi, 2005).
As noted by Guggenberger (2010), examples of the practical application of this procedure are provided by Bloningen (1997) and Hastings (2004).
This preliminary model selection procedure has also been adopted in other areas such as medical statistics,
see e.g. Gardiner et al. (2009) and Mann et al. (2004), and has been implemented in popular statistical computer programmes including SAS, Stata, eViews and R, see
Ajmani (2009, Chapter 7.5.3), Rabe-Hesketh and Skrondal (2012, Chapter 3.7.6), Griffiths et al. (2012, Chapter 10.4) and Croissant and Millo (2008), respectively.

%\noindent \textit{PK: It would be useful to add references that explicitly carry out a two-stage procedure where the first stage is a Hausman test and %the
%second stage is a confidence interval}

So, what is widely used in the analysis of panel data that includes a time-varying covariate, is the following two-stage procedure.
In the first stage, the Hausman pretest is used to decide
whether subsequent inference is made using the random effects model or the fixed effects model (see e.g. Ebbes et al., 2004 and Jackowicz et al., 2013). The second stage is that the inference
of interest is carried out assuming that the model chosen in the first stage had been given to us \textit{a priori}, as the true model.
%This assumption is false and it can lead to an inaccurate inference.
Guggenberger (2010) considers this two-stage procedure when the inference of interest is a hypothesis test about the slope
parameter.
He provides both a local asymptotic analysis of the size of this test and a finite sample analysis (via simulations) of the probability of Type I error. 

In the present paper, we consider the case
that the inference of interest is a confidence interval for the slope parameter.
We prove three new theorems on the finite sample properties of the coverage probability
function of this confidence interval. By the duality between hypothesis tests and confidence intervals,
these new theorems imply corresponding new results when the inference of interest is a hypothesis test (see Remark 1 in Section 5).
Theorem 1 states that the finite sample coverage probability of the confidence interval resulting from the two-stage procedure depends on relatively few parameters. We use Theorem 3 to provide variance reduction by control variates, leading to more efficient simulation-based estimates of
coverage probability. Also, we use Theorem 2 to reduce the time required to compute the minimum coverage probability by a half.
These theorems make it easy to assess, for a wide variety of circumstances, the finite sample effect of the
Hausman pretest on the minimum coverage probability of this confidence interval (or, when the inference of interest is a hypothesis test, the size of this test). 
 We find that the Hausman pretest, with the usual small nominal level of significance, can lead to this confidence interval having minimum coverage probability far below nominal. We also find that if the nominal level of significance is increased to 50\% then the minimum coverage probability is much closer to the nominal coverage
(as shown in Figures 1, 3 and 4).
The results presented in this paper were
computed using programs written in the R programming language, which will be made available in a convenient R package.

%Let $\sigma_{\varepsilon}^2$ and $\sigma_{\mu}^2$ denote the variances of the random error and the random effect, respectively.
In Section 2, we consider the practical situation that the
%parameters $\sigma_{\varepsilon}$ and $\sigma_{\mu}$
random error and random effect variances
are estimated from
the data. We consider three estimators of these variances: the usual unbiased estimators, the maximum likelihood estimators of Hsiao (1986) and the estimators of Wooldridge (2002).
The coverage probability of the confidence interval resulting from the two-stage procedure is determined by 4 known quantities and 5 unknown
parameters. The known quantities are the number of individuals, the number of time points, the nominal significance level of the Hausman pretest and the nominal coverage probability of this confidence interval. The unknown parameters are the random error variance, the random effect variance, the variance of the time-varying covariate, a scalar parameter that determines the correlation matrix of the time-varying covariates and a non-exogeneity
parameter.

If, for given values of the 4 known quantities, we wish to assess the dependence of the coverage probability of the confidence interval resulting from the two-stage procedure on the 5 unknown parameters then we might consider, say, five values for each of these unknown parameters, leading to 3125 parameter combinations. Apart from the daunting task of summarizing so many results,
 it is possible that one might miss important values of the unknown parameters, such as values for which the
coverage probability is particularly low.

Theorem 1 states that, apart from the known quantities, this coverage probability
is actually determined by only 3 unknown
parameters, including the non-exogeneity parameter. If we compute the minimum coverage probability with respect to the non-exogeneity parameter then
we have only 2 unknown parameters and our assessment of the coverage properties of the confidence interval resulting from the two-stage procedure is
greatly simplified. Theorem 2 states that this coverage probability is an even function of the non-exogeneity parameter, so that
the time required to compute this minimum coverage is halved.  We also propose a scaling of the non-exogeneity parameter that takes account of the sample size. In effect, this scaling reduces the number of known quantities that determine this coverage probability
 from 4 to 3.

In Section 3, we consider the coverage probability of the confidence interval
resulting from the two-stage procedure when the random error and random effect variances are assumed to be known.
Theorem 3 states that this coverage probability, conditional on the time-varying covariates, can be found exactly by the evaluation of
the bivariate normal cumulative distribution function. This theorem is important because it is used to reduce the variance of the simulation based estimators of the coverage probability
of the confidence interval resulting from the two-stage procedure (when random error and random effect variances are estimated). As we show in Section 4 this variance reduction is achieved by using control variates.

\medskip

\begin{center}
\textbf{2. THE MODEL AND THE PRACTICAL TWO-STAGE PROCEDURE
(RANDOM ERROR AND RANDOM EFFECT VARIANCES ARE ESTIMATED)}
\end{center}

%\medskip

%\noindent \textit{PK: What word do the econometricians use? %``subject'' or ``individual'' or something else?}

\noindent Let $y_{it}$ and $x_{it}$ denote the response variable and the time-varying covariate, respectively, for the individual $i$ ($i=1, \dots, N$) at time $t$ ($t=1, \dots, T$). Suppose that
\begin{equation}
\label{model}
y_{it} = a + \beta x_{it} + \mu_i + \varepsilon_{it},
\end{equation}
where the $\varepsilon_{it}$'s and the $(\mu_i, x_{i1}, \dots, x_{iT})$'s are independent, the $\varepsilon_{it}$'s are i.i.d. $N(0, \sigma_{\varepsilon}^2)$ and the $\mu_i$'s are i.i.d. $N(0, \sigma_{\mu}^2)$. We call $\beta$ the slope parameter, $\sigma_{\varepsilon}^2$ the error variance
and $\sigma_{\mu}^2$ the random effect variance. Note that the $\varepsilon_{it}$'s and the
$\mu_i$'s are unobserved.
Suppose that the parameter of interest is $\beta$ and that the inference of interest is a confidence interval for $\beta$.
Let $x = (x_{11}, \dots, x_{1T}, x_{21}, \dots, x_{2T}, \dots, x_{N1}, \dots, x_{NT})$.

Also suppose that the $(\mu_i, x_{i1}, \dots, x_{iT})$'s are
i.i.d. multivariate normally distributed with zero mean and covariance matrix
\begin{equation}
\label{CovMatrixMuiXit}
\begin{bmatrix}
\sigma^2_{\mu} & \widetilde{\tau}\sigma_{\mu}\sigma_x e^{\prime} \\
\widetilde{\tau}\sigma_{\mu} \sigma_x e & \sigma^2_x G
\end{bmatrix},
\end{equation}
where $e$ is a $T$ vector of 1's, $G$  is a $T \times T$ matrix with $1$'s on the diagonal and $\tilde{\tau}$ is a parameter that measures the dependence between $\mu_{i}$ and $(x_{i1}, \dots, x_{iT})$. We consider two models for $G$: (a) the off-diagonal elements of $G$ are all $\rho$ (compound symmetry) and (b) the $(i,j)$'th element of $G$ is $\rho^{|i-j|}$
(first order autoregression). We define the ``non-exogeneity parameter'' $\tau$ as follows. For the case of compound symmetry,
$\tau = \widetilde{\tau} \big(T/(1+(T-1)\rho) \big)^{1/2}$ and, for first order autoregression,
$\tau =\widetilde{\tau} \big(( T(1-\rho)+ 2\rho)/(1+\rho) \big)^{1/2}$.
As we show in Appendix A, in both cases $\tau$ is a correlation, so that it lies in the interval $(-1, 1)$.
If $\tau = 0$ then $\mu_i$ and $(x_{i1}, \dots, x_{iT})$ are independent, so that the
$x_{it}$'s are exogenous variables.
Also, as shown in Appendix A, for the compound symmetry case, our definition of the parameter $\tau$ coincides with the definition
given by Guggenberger (2010, p.339) of his parameter $\gamma_1$, which ``measures the degree of failure of the pretest hypothesis''.

 When $\tau = 0$, a confidence interval
for $\beta$ may be found as follows. Assume, initially, that
%$\sigma_{\varepsilon}$ and $\sigma_{\mu}$ are known, so that
$\psi = \sigma_{\mu}  / \sigma_{\varepsilon}$ is known.
Condition on $x$ and use the GLS estimator
$\widehat{\beta}(\psi)$ of $\beta$. Let $z_c = \Phi^{-1}(c)$, where $\Phi$ denotes the standard normal cdf.
 The $1-\alpha$
confidence interval for $\beta$ based on this estimator is
\begin{equation*}
I(\psi) = \left[\widehat{\beta}(\psi) - z_{1-\alpha/2} \big(\text{Var}_0(\widehat{\beta}(\psi) \, | \, x)\big)^{1/2}, \, \widehat{\beta}(\psi)
+ z_{1-\alpha/2} \big(\text{Var}_0(\widehat{\beta}(\psi) \, | \, x)\big)^{1/2} \right],
\end{equation*}
where $\text{Var}_0(\widehat{\beta}(\psi) \, | \, x)$ denotes the variance of $\widehat{\beta}(\psi)$, conditional on $x$ when $\tau = 0$. When $\tau = 0$ the confidence interval $I(\psi)$
has coverage probability $1-\alpha$, conditional on $x$.
Therefore, it has coverage probability $1-\alpha$ unconditionally.
Of course, in practice $\psi$ is unknown and needs to be estimated.
So, in practice, we would use  $I(\widehat{\psi})$,
where $\widehat{\psi}$ is an estimator of $\psi$, as a confidence interval
with nominal coverage probability $1-\alpha$.

If we average \eqref{model} over $t=1, \dots, T$ for each $i=1, \dots, N$ then we obtain
\begin{equation}
\label{average_over_t_model}
\overline{y}_i = a + \beta \overline{x}_i + \mu_i + \overline{\varepsilon}_i,
\end{equation}
where
\begin{equation*}
\overline{y}_i = \frac{1}{T} \sum_{t=1}^T y_{it} \ , \ \ \ \overline{x}_i = \frac{1}{T} \sum_{t=1}^T x_{it}
\ \ \ \text{and} \ \ \ \ \overline{\varepsilon}_i = \frac{1}{T} \sum_{t=1}^T \varepsilon_{it}.
\end{equation*}
This model is called the between effects model.
When $\tau = 0$, an alternative estimator of
$\beta$ is $\widetilde{\beta}_B$, the OLS estimator based on the model \eqref{average_over_t_model}, when we condition on $x$. This estimator does not require a knowledge of $\psi$.

Irrespective of whether $\tau = 0$ or not, we may find a confidence interval for $\beta$ with coverage probability $1-\alpha$ as follows. Subtracting
\eqref{average_over_t_model} from \eqref{model}, we obtain
\begin{equation}
\label{consistent_est_beta_model}
y_{it} - \overline{y}_i = \beta (x_{it} - \overline{x}_i) + (\varepsilon_{it} - \overline{\varepsilon}_i).
\end{equation}
This model is called the fixed effects model. We estimate $\beta$
by $\widetilde{\beta}_W$, the OLS estimator based on this model.
The $1-\alpha$
confidence interval for $\beta$ based on this estimator is
\begin{equation*}
J(\sigma_{\varepsilon}) = \left [\widetilde{\beta}_W - z_{1-\alpha/2} \big(\text{Var}(\widetilde{\beta}_W \, | \, x)\big)^{1/2},
\widetilde{\beta}_W +
z_{1-\alpha/2} \big(\text{Var}(\widetilde{\beta}_W \, | \, x)\big)^{1/2} \right ],
\end{equation*}
where $\text{Var}(\widetilde{\beta}_W \, | \, x)$ denotes
the variance of $\widetilde{\beta}_W$, conditional on $x$.
The confidence interval $J(\sigma_{\varepsilon})$
has coverage probability $1-\alpha$, conditional on $x$.
Therefore, it has coverage probability $1-\alpha$ unconditionally.
Of course, in practice $\sigma_{\varepsilon}$ is unknown and needs to be estimated.
So, in practice, we would use  $J(\widehat{\sigma}_{\varepsilon})$,
where $\widehat{\sigma}_{\varepsilon}$ is an estimator of $\sigma_{\varepsilon}$, as a confidence interval
with nominal coverage probability $1-\alpha$.

In practice, we do not know whether or not $\tau = 0$. As noted in the introduction, the usual procedure is to use a Hausman pretest to test the null hypothesis $\tau = 0$ against the alternative hypothesis $\tau \not= 0$. Assume, for the moment, that $\sigma_{\varepsilon}$ and $\sigma_{\mu}$ are known. We consider this pretest, based on the test statistic
\begin{equation}
\label{Def_H}
H(\sigma_{\varepsilon},\sigma_{\mu}) =  \frac{(\widetilde{\beta}_W-\widetilde{\beta}_B)^2}{\text{Var}(\widetilde{\beta}_W \, | \, x) + \text{Var}_0(\widetilde{\beta}_B \, | \, x)},
\end{equation}
where $\text{Var}_0(\widetilde{\beta}_B \, | \, x)$ denotes the variance of
$\widetilde{\beta}_B$ conditional on $x$ and assuming that $\tau = 0$.
This test statistic has a $\chi^2_1$ distribution under the null hypothesis $\tau = 0$, conditional on $x$. Therefore, this test statistic has this distribution under this null hypothesis, unconditionally.  Suppose that we accept the null hypothesis $\tau = 0$ if
$H(\sigma_{\varepsilon},\sigma_{\mu})
\le z^2_{1 - \widetilde{\alpha}/2}$; otherwise we reject this null hypothesis.
Note that $\widetilde{\alpha}$ is the level of significance of this test, conditional on $x$,
assuming that $\sigma_{\varepsilon}$ and $\sigma_{\mu}$ are known.

We now describe the two-stage procedure assuming, for the moment, that $\sigma_{\varepsilon}$ and $\sigma_{\mu}$ are known.
If the null hypothesis $\tau=0$ is accepted then we use the confidence interval $I(\psi)$; otherwise we use the confidence interval $J(\sigma_{\varepsilon})$. Let $K(\sigma_{\varepsilon},\sigma_{\mu})$ denote the confidence interval, with nominal coverage
$1-\alpha$, that results from this two-stage procedure. Of course, in practice, $\sigma_{\varepsilon}$ and $\sigma_{\mu}$ are not known and need to be estimated. So, in practice, the two-stage procedure results in the confidence interval $K(\widehat{\sigma}_{\varepsilon},\widehat{\sigma}_{\mu})$ where $\widehat{\sigma}_{\varepsilon}$ and $\widehat{\sigma}_{\mu}$ denote estimators of $\sigma_{\varepsilon}$ and $\sigma_{\mu}$, respectively.  We consider the usual unbiased estimators, maximum likelihood estimators and Wooldridge's estimators (described in Appendix B).  The unconditional coverage probability of the confidence interval constructed from this two-stage procedure is denoted $P(\beta \in K(\widehat{\sigma}_{\varepsilon}, \widehat{\sigma}_{\mu} ))$.

As stated in the introduction, $P(\beta \in K(\widehat{\sigma}_{\varepsilon}, \widehat{\sigma}_{\mu} ))$
is determined by the 4 known quantities $N$, $T$, $\widetilde{\alpha}$ and $1-\alpha$ and the 5 unknown parameters
 $\sigma_{\varepsilon}^2$, $\sigma_{\mu}^2$, $\sigma_x^2$, $\rho$ and $\tau$. As explained in the introduction,
 the assessment of the dependence of this probability on these 4 known quantities and 5 unknown parameters is a daunting task that could
 easily miss important parameter values. However,
 the following theorem shows that, for given values of the known quantities, this probability depends on the 5 unknown
 parameters only through the 3 unknown parameters $\psi = \sigma_{\mu}/\sigma_{\varepsilon}$, $\rho$ and $\tau$.
 Thus, for fixed $N$, $T$, $\widetilde{\alpha}$ and $1-\alpha$, we only need to consider what happens as $\psi$, $\rho$ and $\tau$ are varied
  instead of what happens as $\sigma_{\varepsilon}$, $\sigma_{\mu}$, $\sigma_x$, $\rho$ and $\tau$ are varied.
 If we compute the minimum over $\tau$ of $P(\beta \in K(\widehat{\sigma}_{\varepsilon}, \widehat{\sigma}_{\mu} ))$
 then we are left with only 2 unknown parameters $\psi$ and $\rho$.

%\newpage

\begin{theorem}
\label{thm: CovProb_DependOnPsi}

%For given values of $\tau$, $\alpha$, $\widetilde{\alpha}$, $\sigma_x$, $\sigma_{\varepsilon}$ and $\sigma_{\mu}$ the coverage probability of $K(\sigma_{\varepsilon}, \sigma_{\mu})$ (\textit{PK: we need to have $\sigma_x$ somewhere in here. We need to keep in mind that for the compound symmetry case there is no restriction on the value of $\tau$, whereas for the first order autoregression there is a restriction}) does not depend on $\sigma_x$ and only depends on $\sigma_{\varepsilon}$ and $\sigma_{\mu}$ through the ratio $\psi$.

\smallskip

\noindent For $(\widehat{\sigma}_{\varepsilon},\widehat{\sigma}_{\mu})$ any of the
pairs of estimators listed in Appendix B, the unconditional coverage probability $P(\beta \in K(\widehat{\sigma}_{\varepsilon}, \widehat{\sigma}_{\mu} ))$ is determined by $N$ (the number of individuals), $T$ (the number of time points), $\widetilde{\alpha}$ (the nominal significance level of the Hausman pretest), $1 - \alpha$ (the nominal coverage probability), $\psi$ (the ratio $\sigma_{\mu}/\sigma_{\varepsilon}$), $\rho$ (the parameter that determines $G$) and $\tau$ (the non-exogeneity parameter).  Given these quantities, the coverage probability does not depend on either $\sigma^2_{\varepsilon}$ (the variance of the random error) or $\sigma^2_{\mu}$ (the variance of the random effect) or $\sigma^2_x$ (the variance of the time-varying covariate $x_{it}$).

\end{theorem}

The proof of Theorem 1 is provided in Appendix C.
We use simulations to compute $P(\beta \in K(\widehat{\sigma}_{\varepsilon}, \widehat{\sigma}_{\mu} ))$, employing
 variance reduction by control variates, as described in Section 4.  When we compute this coverage probability, we also make use of the following theorem.

\begin{theorem}
\label{thm: eveness}
Suppose that $N$, $T$, $\widetilde{\alpha}$, $1-\alpha$, $\psi$ and $\rho$ are fixed.
When $\sigma_{\varepsilon}$ and  $\sigma_{\mu}$ are replaced by any of the pairs of estimators listed in Appendix B,
the unconditional coverage probability $P(\beta \in K(\widehat{\sigma}_{\varepsilon}, \widehat{\sigma}_{\mu} ))$
is an even function of $\tau \in (-1,1)$.
\end{theorem}

The proof of Theorem \ref{thm: eveness} is provided in Appendix C.
A remarkable feature of the proofs of both Theorems 1 and 2 is that they are carried out without relying on a simple expression for the
coverage probability $P(\beta \in K(\widehat{\sigma}_{\varepsilon}, \widehat{\sigma}_{\mu} ))$.
Using Theorem \ref{thm: eveness}, we only need to consider $\tau$ in the interval $[0, 1)$, which means that we have reduced the number of simulations needed to estimate the coverage probability function (or its minimum) by half.
By Remark 2 of Section 5,
we expect this coverage probability, considered as a function of $\lambda = N^{1/2} \tau$,
to have a stable shape when $N$ is varied over a wide range of medium to large values.
We therefore plot this coverage probability as a function of $\lambda$, instead of $\tau$.
Of course,
the set of possible values of $\lambda$ changes with $N$, since $\tau \in (-1, 1)$. In other words,
$\lambda \in (-N^{1/2}, N^{1/2})$.

We now examine the influence that the nominal level of significance $\widetilde{\alpha}$ of the Hausman pretest has on the coverage probability function $P(\beta \in K(\widehat{\sigma}_{\varepsilon}, \widehat{\sigma}_{\mu} ))$.
Suppose that
$\widehat{\sigma}_{\varepsilon}$ and $\widehat{\sigma}_{\mu}$
are the usual unbiased estimators of $\sigma_{\varepsilon}$ and $\sigma_{\mu}$, respectively (described in Appendix B).
Consider the case that the matrix $G$ has off-diagonal
 elements $\rho$ (compound symmetry), where $\rho = 0.3$, $N = 100$, $T = 3$, $\psi =  \sigma_{\mu} / \sigma_{\varepsilon} = 1/3$ and the
nominal coverage probability $1-\alpha = 0.95$.
In practice, it is common to use a small value of $\widetilde{\alpha}$, such as 0.05 or 0.01.
As noted by Guggenberger (2010), examples of practical applications that have used a small $\widetilde{\alpha}$ for the Hausman pretest are provided by Gaynor et al. (2005, p.245) and Bedard and Deschenes (2006, p.189).
Figure 1 presents graphs of the coverage probability
$P(\beta \in K(\widehat{\sigma}_{\varepsilon}, \widehat{\sigma}_{\mu} ))$,
considered as a function of $\lambda = N^{1/2} \tau$.
Each graph is computed using the variance reduction method and the common random numbers (to produce smoother graphs)
described in Section 4.
The number of simulation runs used to compute each graph is $M = 20000$.
The bottom (solid line) graph is for nominal significance level $\widetilde{\alpha} = 0.05$ of the Hausman pretest.
This graph falls well below the nominal
coverage for a wide interval of values of $\lambda$, with the minimum of the coverage probability approximately equal to 0.75.
Suppose that we choose the significance level of the Hausman pretest to be quite large, say $\widetilde{\alpha}=0.50$.  Now the Hausman pretest is  more likely to reject the null hypothesis that $\tau =0$ and therefore more likely to choose the fixed effects model for the construction of the
confidence interval.
The middle (dashed line) graph is for nominal significance level $\widetilde{\alpha} = 0.50$ of the Hausman pretest.
Although this graph is still below the nominal coverage, there has been a large improvement.
Similar graphs, for $\widetilde{\alpha} = 0.05$ and $\widetilde{\alpha} = 0.50$, are obtained when $G$ has $(i,j)$'th element $\rho^{|i-j|}$ (first order autoregression) and  $\rho = 0.36$.
\begin{center}
\textit{Figure 1 near here}
\end{center}
As noted earlier, for given values of $T$, $\widetilde{\alpha}$, $1-\alpha$, $\psi$ and $\rho$, we expect the graph of the coverage probability $P(\beta \in K(\widehat{\sigma}_{\varepsilon}, \widehat{\sigma}_{\mu}))$, expressed as a function of $\lambda$, to have a stable shape when $N$ is varied over a wide range of medium to large values.
Suppose that
$\widehat{\sigma}_{\varepsilon}$ and $\widehat{\sigma}_{\mu}$
are the usual unbiased estimators of $\sigma_{\varepsilon}$ and $\sigma_{\mu}$, respectively.
Consider the case that the matrix $G$ has off-diagonal
 elements $\rho$ (compound symmetry), where $\rho = 0.4$,  $T = 5$, $\psi =  \sigma_{\mu} / \sigma_{\varepsilon} = 1/2$ and the
nominal coverage probability $1-\alpha = 0.95$.
Figure 2 presents graphs of the coverage probability
$P(\beta \in K(\widehat{\sigma}_{\varepsilon}, \widehat{\sigma}_{\mu} ))$,
considered as a function of $\lambda = N^{1/2} \tau$, for $N = 25, 50, 100$ and 1000.
Each graph is computed using the variance reduction method and the common random numbers
described in Section 4.
The number of simulation runs used to compute each graph is $M = 5000$.
The value of $\lambda$ for given $N$ must be less than $N^{1/2}$.  So when $N = 25$, $\lambda$ must be less than 5 and when $N = 50$, $\lambda$ must be less than $50^{1/2}$.  This is why the graphs of the coverage probability for these values of $N$ end before $\lambda = 8$.
These graphs do, indeed, have the expected stable shape for $N = 25, 50, 100$ and 1000.
\begin{center}
\textit{Figure 2 near here}
\end{center}
As noted in the Introduction and in Section 2, if we compute the minimum over $\tau$ of the coverage probability
$P(\beta \in K(\widehat{\sigma}_{\varepsilon}, \widehat{\sigma}_{\mu}))$ then we are left with only two unknown parameters, $\psi$ and $\rho$.
%For each $\tau \in [0, 1)$, this coverage probability can be estimated using the simulation methods described in this section.
If we fix $\psi$ then the minimum coverage depends only on $\rho$, where $\rho \in (-1, 1)$, as it is a correlation.
Suppose that
$\widehat{\sigma}_{\varepsilon}$ and $\widehat{\sigma}_{\mu}$
are the usual unbiased estimators of $\sigma_{\varepsilon}$ and $\sigma_{\mu}$, respectively.
Consider the cases that the matrix $G$ has (a) off-diagonal
 elements $\rho$ (compound symmetry) and (b) $(i,j)$'th element $\rho^{|i-j|}$ (first order autoregression).
Suppose that $N = 100$, $T = 3$, $\psi =  \sigma_{\mu} / \sigma_{\varepsilon} = 1/3$ and the
nominal coverage probability $1-\alpha = 0.95$.
Figure 3 presents graphs of the coverage probability
$P(\beta \in K(\widehat{\sigma}_{\varepsilon}, \widehat{\sigma}_{\mu} ))$, minimized over $\tau$,
considered as a function of $\rho$.
Each estimate of the minimum coverage is found using the common random numbers and, for compound symmetry,
 the variance reduction method described in Section 4.
Similarly to Figure 1, we see a vast improvement in the minimum coverage by letting $\widetilde{\alpha} = 0.50$ rather than choosing $\widetilde{\alpha}$ to be the commonly used, smaller value 0.05.
\vspace{-0.5cm}
\begin{center}
\textit{Figure 3 near here}
\end{center}
In practice, $\psi$ is not known and must be estimated from the data.  However, one is likely to have some background knowledge about $\rho$.  This suggests that we fix $\rho$ and plot the graph of the coverage probability
$P(\beta \in K(\widehat{\sigma}_{\varepsilon}, \widehat{\sigma}_{\mu} ))$, minimized over $\tau$, as a function of $\psi$.
Suppose that
$\widehat{\sigma}_{\varepsilon}$ and $\widehat{\sigma}_{\mu}$
are the usual unbiased estimators of $\sigma_{\varepsilon}$ and $\sigma_{\mu}$, respectively.
Consider the cases that the matrix $G$ has (a) off-diagonal
 elements $\rho$ (compound symmetry) and (b) $(i,j)$'th element $\rho^{|i-j|}$ (first order autoregression),
 where $\rho = 0.4$.
Suppose that $N = 100$, $T = 3$, and the nominal
 coverage probability $1-\alpha = 0.95$.
Figure 4 presents graphs of the coverage probability
$P(\beta \in K(\widehat{\sigma}_{\varepsilon}, \widehat{\sigma}_{\mu} ))$, minimized over $\tau$,
considered as a function of $\psi$.
For nominal significance level $\widetilde{\alpha} = 0.05$ of the Hausman pretest, this minimized coverage probability
is far below the nominal coverage
for $\psi$ approximately equal to 0.2. However, for nominal significance level $\widetilde{\alpha} = 0.5$ of the Hausman pretest, we see
(once more) a dramatic
improvement in the minimum coverage probability.
\begin{center}
\textit{Figure 4 near here}
\end{center}

%\bigskip
%\newpage
\begin{center}
\textbf{3. THE TWO-STAGE PROCEDURE WHEN RANDOM ERROR AND RANDOM EFFECT VARIANCES ARE ASSUMED KNOWN}
\end{center}

%\medskip

\noindent In this section we suppose that $\sigma_{\varepsilon}$ and $\sigma_{\mu}$ are known.
In this case, the confidence interval resulting from the two-stage procedure is denoted by $K(\sigma_{\varepsilon}, \sigma_{\mu})$.
We also suppose that the matrix $G$, which appears in the expression \eqref{CovMatrixMuiXit}
for the covariance matrix of $(\mu_i, x_{i1}, \dots, x_{iT})$,
 has 1's on the diagonal and $\rho$ elsewhere (compound symmetry). We show that the coverage probability of this confidence interval, conditional on $x$,  can be computed exactly using the bivariate normal distribution. We employ this computed value in Section 4 to
 find a control variate that is used for variance reduction for the estimation by simulation of
%the unconditional coverage probability of the confidence interval resulting from the two-stage procedure
$P(\beta \in K(\widehat{\sigma}_{\varepsilon}, \widehat{\sigma}_{\mu} ))$, when $\sigma_{\varepsilon}$ and $\sigma_{\mu}$ are unknown.

Let $P\big(\beta \in K(\sigma_{\varepsilon}, \sigma_{\mu}) \, \big| \, x \big)$ denote the coverage probability of
$K(\sigma_{\varepsilon}, \sigma_{\mu})$, conditional on $x$.
Observe that $P\big(\beta \in K(\sigma_{\varepsilon}, \sigma_{\mu}) \, \big| \, x \big)$ is equal to
\begin{align}
\label{CondCPKnown1}
\notag
&P \Big(\beta \in I(\sigma_{\varepsilon},\sigma_{\mu}), \, H(\sigma_{\varepsilon},\sigma_{\mu}) \le z_{1-\widetilde{\alpha}/2}^2 \, \Big| \, x \Big)
+ P \Big(\beta \in J(\sigma_{\varepsilon}), \, H(\sigma_{\varepsilon},\sigma_{\mu}) > z_{1-\widetilde{\alpha}/2}^2 \, \Big| \, x \Big) \\
&= P \big ( |g_I| \le z_{1-\alpha/2}, \, |h| \le z_{1-\widetilde{\alpha}/2} \, \big| \, x \big)
+ P \big ( |g_J| \le z_{1-\alpha/2}, \, |h| > z_{1-\widetilde{\alpha}/2} \, \big| \, x \big),
\end{align}
where
$g_I = \big(\widehat{\beta}(\psi)-\beta \big)/\big(\text{Var}_0(\widehat{\beta}(\psi)|x) \big)^{1/2}$,
$g_J = (\widetilde{\beta}_W-\beta)/\big(\text{Var}(\widetilde{\beta}_W|x)\big)^{1/2}$ and
$h = (\widetilde{\beta}_W - \widetilde{\beta}_B)/\big(\text{Var}(\widetilde{\beta}_W|x) + \text{Var}_0(\widetilde{\beta}_B|x)\big)^{1/2}$.
By the law of total probability, \eqref{CondCPKnown1}
is equal to the sum of $(1-\alpha)$ and
\begin{equation}
\label{CondCPKnown2}
P \big ( |g_I| \le z_{1-\alpha/2}, \, |h| \le z_{1-\widetilde{\alpha}/2} \, \big| \, x \big)
- P \big ( |g_J| \le z_{1-\alpha/2}, \, |h| \le z_{1-\widetilde{\alpha}/2} \, \big| \, x \big).
\end{equation}
The first and second terms in this expression are determined by the conditional
distributions of the random vectors $(g_I, h)$ and $(g_J, h)$,
respectively. Theorem 3 gives these distributions, whose description requires the introduction of
the following notation.
Let $\overline{x} = (NT)^{-1}\sum_{i=1}^N\sum_{t=1}^Tx_{it}$, $\SSB = \sum_{i=1}^N(\overline{x}_i - \overline{x})^2$ (``sum of squares between") and $\SSW = \sum_{i=1}^N \sum_{t=1}^T (x_{it} - \overline{x}_i)^2$ (``sum of squares within").
We define $p^2(x)$ to be $\SSB/\text{Var}(\overline{x}_i)$, where $\text{Var}(\overline{x}_i)$ is given in Appendix A.
Also let $r(x) = \SSB/\SSW$ and $q(\psi, T) = \psi^2 + (1/T)$.
The following theorem is proved in Appendix C.

\begin{theorem}
\label{known coverage}

Conditional on $x$, $(g_I, h)$ and $(g_J, h)$ have bivariate normal distributions, where $E(g_J \, | \, x) =  0$,
${\rm Var}(g_J \, | \, x) = 1$,
\begin{align*}
&E(g_I \, | \, x) =  \displaystyle{\frac{\tau \psi p(x)}{\big(q(\psi, T) + q^2(\psi, T)/r(x)\big)^{1/2}}}, \;
{\rm Var}(g_I \, | \, x) = 1 - \displaystyle{\frac{\tau^2 \psi^2}{q(\psi, T)  + q^2(\psi, T) / r(x)}}, \\
\\
%&E(g_J \, | \, x) =  0,
%{\rm Var}(g_J \, | \, x) = 1, \\
%\\
&E(h \, | \, x) =  \displaystyle{\frac{-\tau \psi p(x)}{(r(x) + q(\psi,T))^{1/2}}}, \;
{\rm Var}(h \, | \, x) = 1 - \displaystyle{\frac{\tau^2 \psi^2 }{r(x) + q(\psi,T)}}, %\\
%\\
\end{align*}
\begin{align*}
&{\rm Cov}(g_I, h \, | \, x) = \displaystyle{\frac{\tau^2 \psi^2}{\big(q(\psi,T) r(x)+ q^2(\psi,T)\big)^{1/2} \big(1 + q(\psi,T)/r(x)\big)^{1/2}}} \\
\\
&\text{and} \ \ \ {\rm Cov}(g_J, h \, | \, x) = \displaystyle{\frac{1}{\big(1 + q(\psi,T)/r(x) \big)^{1/2}}}.
\end{align*}

\end{theorem}

\smallskip

Thus, when $\sigma_{\varepsilon}$ and $\sigma_{\mu}$ are known,
$P \big(\beta \in K(\sigma_{\varepsilon}, \sigma_{\mu})\, \big| \, x \big)$
can be found easily by evaluation of the bivariate normal cumulative distribution function
in the expression \eqref{CondCPKnown2}.
Similarly to Theorem 1,
this probability
is determined by $N$ (the number of individuals), $T$ (the number of time points), $x$ (the vector of time-varying covariates), $\widetilde{\alpha}$ (the nominal significance level of the Hausman pretest), $1-\alpha$ (the nominal coverage probability), $\psi$ (the ratio $\sigma_{\mu}/\sigma_{\varepsilon}$), $\rho$ (the parameter that determines $G$) and $\tau$ (the non-exogeneity parameter).
Note that the dependence on $\rho$ is through $p(x)$.
Also, similarly to Theorem 2, $P(\beta \in K(\sigma_{\varepsilon}, \sigma_{\mu})|x)$ is an even function of $\tau \in (-1,1)$.  These results may be proved using similar, but much simpler, arguments to those used in the proofs of Theorems 1 and 2.

\begin{center}
\textbf{4. SIMULATION METHODS, INCLUDING THE USE OF VARIANCE REDUCTION, WHEN THE RANDOM ERROR AND RANDOM EFFECT VARIANCES ARE UNKNOWN}
\end{center}
%\medskip

\noindent In Section 3 we described how to find the coverage probability of the confidence interval resulting from the two-stage procedure,
conditional on $x$ when $\sigma_{\varepsilon}$ and $\sigma_{\mu}$ are known, using the bivariate normal distribution.  In the practically important case that $\sigma_{\varepsilon}$ and $\sigma_{\mu}$ are replaced by estimators, we can no longer use the bivariate normal distribution to find this coverage probability.  Instead we estimate the coverage probability using a simulation consisting of $M$ independent simulation runs.  We consider the model \eqref{model} and choose the intercept $a = 0$, the parameter of interest $\beta = 0$ and the
values for  $N$ (the number of individuals), $T$ (the number of time points), $\widetilde{\alpha}$ (the nominal significance level of the Hausman pretest), $1 -\alpha$ (the nominal coverage probability), $\sigma^2_{\varepsilon}$ (the variance of the random error), $\sigma^2_{\mu}$ (the variance of the random effect) and $\sigma^2_{x}$ (the variance of the covariate $x_{it}$). Of course, by Theorem 1, the coverage probability
does not depend on either $a$, $\beta$ or $\sigma^2_{x}$ and depends on $\sigma^2_{\varepsilon}$ and $\sigma^2_{\mu}$ only through
$\psi=\sigma_{\mu}/\sigma_{\varepsilon}$. The simulation methods described in this section apply to any of the pairs of estimators $\widehat{\sigma}_{\varepsilon}$ and $\widehat{\sigma}_{\mu}$ listed in Appendix B.

On the $k$'th simulation run, we generate observations of the $\varepsilon_{it}$'s and $(\mu_i, x_{i1}, \dots, x_{iT})$'s using the assumptions made in Section 2, i.e. the $\varepsilon_{it}$'s are i.i.d. $N(0, \sigma^2_{\varepsilon})$ and the $(\mu_i, x_{i1}, \dots, x_{iT})$'s are i.i.d. with a multivariate normal distribution with mean $0$ and covariance matrix \eqref{CovMatrixMuiXit}.
 Let $x^k$ denote the observed value of $x$ for this run.
For the observed values in this simulation run,
we compute the following three quantities.
The confidence interval resulting from the two-stage procedure, when $\sigma_{\varepsilon}$ and $\sigma_{\mu}$ are assumed known, is denoted by $K_k(\sigma_{\varepsilon}, \sigma_{\mu})$.
The confidence interval resulting from the two-stage procedure,
when $\sigma_{\varepsilon}$ and $\sigma_{\mu}$ are estimated by $\widehat{\sigma}_{\varepsilon}$ and $\widehat{\sigma}_{\mu}$,
respectively, is denoted by $K_k(\widehat{\sigma}_{\varepsilon}, \widehat{\sigma}_{\mu})$.
The coverage probability of $K(\sigma_{\varepsilon}, \sigma_{\mu})$, conditional on $x^k$, when $\sigma_{\varepsilon}$ and $\sigma_{\mu}$ are assumed known, is
$P \big(\beta \in K(\sigma_{\varepsilon}, \sigma_{\mu}) \, \big| \, x^k \big)$. Note that this conditional coverage probability is computed exactly using the bivariate normal distributions given in Theorem 3.

Let $\CP = P(\beta \in K(\widehat{\sigma}_{\varepsilon}, \widehat{\sigma}_{\mu}))$, the coverage probability of $K(\widehat{\sigma}_{\varepsilon}, \widehat{\sigma}_{\mu})$.
We use the notation
\begin{equation*}
{\cal I}({\cal A}) =
\begin{cases}
1 &\text{if } {\cal A} \ \ \text{is true} \\
0 &\text{if } {\cal A} \ \ \text{is false}
\end{cases}
\end{equation*}
where ${\cal A}$ is an arbitrary statement.
Now define the unbiased estimator
\begin{equation*}
\widehat{\CP} = \frac{1}{M} \sum_{k=1}^M \mathcal{I} \big(\beta \in K_k(\widehat{\sigma}_{\varepsilon}, \widehat{\sigma}_{\mu}) \big)
\end{equation*}
of \CP. This is the usual ``brute-force'' simulation estimator of $\CP$.
We estimate the variance of this estimator by noting that it is a binomial proportion.
%Since this estimator is a binomial proportion, $\sqrt{\widehat{\CP}(1-\widehat{\CP})/M}$
%is an estimate of its standard deviation.
Let $\CPK = P \big(\beta \in K(\sigma_{\varepsilon}, \sigma_{\mu}) \big)$, the coverage probability of $K(\sigma_{\varepsilon}, \sigma_{\mu})$, when $\sigma_{\varepsilon}$ and $\sigma_{\mu}$ are assumed known. Now define the unbiased estimator
\begin{equation*}
\widehat{\CPK} = \frac{1}{M} \sum_{k=1}^M \mathcal{I} \big(\beta \in K_k(\sigma_{\varepsilon}, \sigma_{\mu}) \big)
\end{equation*}
of \CPK.
By the \textsl{double expectation theorem}, $\CPK = E_x \big(P \big(\beta \in K(\sigma_{\varepsilon}, \sigma_{\mu}) \, \big| \, x \big) \big)$.
Thus another unbiased estimator of $\CPK = P(\beta \in K(\sigma_{\varepsilon}, \sigma_{\mu}))$ is
\begin{equation*}
\widetilde{\CPK} = \frac{1}{M} \sum_{k=1}^M P \big(\beta \in K(\sigma_{\varepsilon}, \sigma_{\mu}) \, \big| \, x^k \big),
\end{equation*}
which is a much more accurate estimator of $\CPK$  than $\widehat{\CPK}$.

Define the \textsl{control variate} $\widehat{\CPK} - \widetilde{\CPK}$, which has expected value zero. The simulation-based unbiased estimator of
$\CP = P \big(\beta \in K(\widehat{\sigma}_{\varepsilon}, \widehat{\sigma}_{\mu}) \big)$
that employs variance reduction using this control variate,
is
\begin{equation*}
\widetilde{\CP} = \widehat{\CP} - \left (\widehat{\CPK} - \widetilde{\CPK} \right).
\end{equation*}
We expect that the correlation between $\widehat{\CP}$ and $\widehat{\CPK}$ will be close to 1.
Since $\widetilde{\CPK}$ is a much more accurate estimator of $\CPK$  than $\widehat{\CPK}$,
we expect that the correlation between $\widehat{\CP}$ and the control variate $\widehat{\CPK} - \widetilde{\CPK}$
will also be close to 1. Note that
\begin{equation*}
\widetilde{\CP}
= \frac{1}{M} \sum_{k=1}^M
\Big( \mathcal{I}\big(\beta \in K_k(\widehat{\sigma}_{\varepsilon}, \widehat{\sigma}_{\mu}) \big)
-  \mathcal{I}\big(\beta \in K_k(\sigma_{\varepsilon}, \sigma_{\mu}) \big)
+ P\big(\beta \in K(\sigma_{\varepsilon}, \sigma_{\mu}) \, \big| \, x^k \big) \Big).
\end{equation*}
We estimate the variance of this estimator by noting that it is an average of i.i.d. random variables.

We evaluate the efficiency gain of using $\widetilde{\CP}$ to estimate the coverage probability $\CP$ over $\widehat{\CP}$,
as follows.  Let $\widehat{\T}$ and $\widetilde{\T}$ denote the times taken to carry out $M$ simulation runs when we estimate $\CP$ by
$\widehat{\CP}$ and $\widetilde{\CP}$, respectively.
The efficiency
%($RE$)
of the control variate estimator $\widetilde{\CP}$ relative to the ``brute-force'' estimator $\widehat{\CP}$ is
\begin{equation*}
\frac{\widehat{\T}}{\widetilde{\T}} \frac{\text{Var}(\widehat{\CP})}{\text{Var}(\widetilde{\CP})}.
\end{equation*}
The larger this relative efficiency, the greater the gain in using the control variate estimator $\widetilde{\CP}$, by comparison with using the ``brute-force'' estimator $\widehat{\CP}$.
To give an example of the efficiency gained by using $\widetilde{\CP}$ compared to $\widehat{\CP}$, when estimating $\CP$, we set $G=I$ (the $T\times T$ identity matrix), $N= 100$, $T=3$, $\tau = 0$, $\psi = 1/3$, $\alpha = \widetilde{\alpha} = 0.05$ and number of simulation runs $M = 10,000$.  We obtain $\widehat{\T} = 179.37$ seconds, $\widetilde{T} = 211.51$ seconds, $\text{Var}(\widehat{\CP}) = 5.613591\times10^{-6}$ and $\text{Var}(\widetilde{\CP}) = 1.39591\times 10^{-6}$.  The time ratio is $\widehat{\T}/\widetilde{\T} = 0.848045$ and the variance ratio is $\text{Var}(\widehat{\CP})/\text{Var}(\widetilde{\CP}) = 4.92597$, so the efficiency of $\widetilde{\CP}$ relative to $\widehat{\CP}$ is approximately 4.17.  In other words, it would take approximately 4.17 times as long to compute the ``brute-force" estimator with the same accuracy as the control variate estimator.

We also use \textsl{common random numbers} to create smoother plots of the estimated coverage probability, as a function of $\lambda$.
The estimates of the coverage probability are computed for an equally-spaced grid of values of $\lambda$.
On the $k$'th simulation run we generate an observation of $(\mu_i, x_{i1}, \dots, x_{iT})$ by linearly transforming observations of $T+1$ independent
$N(0,1)$ random numbers.
So, on the $k$'th simulation run, for each value of $\lambda$ in the grid, we use the same random numbers that are used to generate the observations of the $\varepsilon_{it}$'s and
the $(\mu_i, x_{i1}, \dots, x_{iT})$'s.  These observations are then used to construct our simulation-based estimate of $\CP$. Therefore on the $k$'th simulation run, for each value of $\lambda$, we have an estimate of the coverage probability using the same random numbers.

%\newpage
%\bigskip
\begin{center}
\textbf{5. REMARKS}
\end{center}

%\smallskip

\noindent {\sl Remark 1:} \ As one would expect from the duality between hypothesis tests and confidence intervals, our results have important implications for the actual size of a hypothesis test for the slope parameter,
with nominal significance level $\alpha$, following a Hausman pretest,
when the random error and random effect variances are estimated by one of the pairs of estimators described in Appendix B. Consider the following two-stage
procedure. As previously, in the first stage we test the null hypothesis $\tau = 0$ against the alternative hypothesis $\tau \ne 0$
as follows.
We accept this null hypothesis if
$H(\widehat{\sigma}_{\varepsilon},\widehat{\sigma}_{\mu})
\le z^2_{1 - \widetilde{\alpha}/2}$; otherwise we reject this null hypothesis.
In the second stage, we test the null hypothesis $H_0: \beta = \beta_0$ against the alternative hypothesis $H_1: \beta \ne \beta_0$ as follows.
If the null hypothesis $\tau = 0$ has been accepted then we test $H_0: \beta = \beta_0$ against $H_1: \beta \ne \beta_0$ at the nominal significance level $\alpha$, using the test statistic $( \widehat{\beta}(\widehat{\psi}) - \beta_0) / ( \widehat{\text{Var}}_0(\widehat{\beta}(\psi) \, | \, x))^{1/2}$, which has a nominal $N(0,1)$
distribution under $H_0$. If, on the other hand, the null hypothesis $\tau = 0$ has been rejected then we test
$H_0: \beta = \beta_0$ against $H_1: \beta \ne \beta_0$ at the nominal significance level $\alpha$, using the test statistic $(\widetilde{\beta}_W - \beta_0) / (\widehat{\text{Var}}(\widetilde{\beta}_W \, | \, x))^{1/2}$, which has a nominal $N(0,1)$ distribution under $H_0$.

It may be shown that the probability of rejecting
the null hypothesis $H_0: \beta = \beta_0$, when it is true, is
$1 - P(\beta_0 \in K(\widehat{\sigma}_{\varepsilon}, \widehat{\sigma}_{\mu}))$. By Theorem 1, this probability of rejection is determined
by $N$, $T$, $\widetilde{\alpha}$, $1-\alpha$, $\psi$, $\rho$ and $\tau$.
In other words, this probability of rejection does not depend on $\sigma_x^2$ and depends on $\sigma_{\varepsilon}^2$ and $\sigma_{\mu}^2$ only through $\psi = \sigma_{\mu} / \sigma_{\varepsilon}$. Since $P(\beta_0 \in K(\widehat{\sigma}_{\varepsilon}, \widehat{\sigma}_{\mu}))$ does not depend on either $a$ or $\beta_0$, we may (without loss of
generality) suppose that $a = 0$ and $\beta_0 = 0$. Also, by Theorem 2, this probability of rejection is an even function of $\tau$ for fixed
$N$, $T$, $\widetilde{\alpha}$, $1-\alpha$, $\psi$ and $\rho$. Hence,
for fixed
$N$, $T$, $\widetilde{\alpha}$, $1-\alpha$, $\psi$ and $\rho$, the size of the hypothesis test of
$H_0: \beta = \beta_0$ against
$H_1: \beta \ne \beta_0$ is equal to
\begin{equation*}
1 - \inf_{\tau \in [0,1)} P \big(\beta_0 \in K(\widehat{\sigma}_{\varepsilon}, \widehat{\sigma}_{\mu}) \big),
\end{equation*}
which can be computed efficiently using our methods.

\medskip

\noindent {\sl Remark 2:} \
Crossover trials are widely used in medicine and pharmaceutics. For many years, the following two-stage procedure was the standard
method for the analysis of data from a two-treatment two-period (AB/BA) crossover trial. In the first stage, a test of the null hypothesis of
zero differential carryover is used to decide whether subsequent inference is made using all of the data or only the data from the
first period (which is unaffected by carryover). In the second stage, we carry out the inference of interest assuming that the model
chosen in the first stage had been given to us \textit{a priori}, as the true model. In a landmark paper, Freeman (1989) considers the case
that the inference of interest is a confidence interval for the difference of the effects of the two treatments. For simplicity,
he assumes that the error and subject variances are known and derives a formula for the coverage probability that
has some similarity to the formula \eqref{CondCPKnown1} of Section 3, in that both of these formulas are evaluated using the bivariate normal distribution. Freeman's
conclusion is that the two-stage procedure for crossover trials ``is too potentially misleading to be of practical use''.

Our choice of the scaling $\lambda = N^{1/2} \tau$ is motivated by Freeman's (1989) scaling of the differential carryover by the square root of the sample size.  We expect that the coverage probability, considered as a function of $\lambda$, for given $N$ will be reflective of the coverage probability function as $N \rightarrow \infty$.  This expectation is verified by Figure 2.

\smallskip

\noindent {\sl Remark 3:} \
The Hausman pretest is an example of preliminary statistical (i.e. data based) model selection. Other examples include model selection by minimizing a criterion such as the Akaike Information Criterion or the Bayesian Information Criterion. The effects of preliminary statistical model selection on confidence intervals can range from the benign to the very harmful, depending on the class of models under consideration, the known aspects of the model, the parameter of interest and the model selection procedure employed (Kabaila, 1995, 2009 and Kabaila and Leeb, 2006). In other words, each case needs to be considered individually on its merits.

%\bigskip
%\medskip
\newpage

\begin{center}
\textbf{6. CONCLUSION}
\end{center}

%\medskip
%\smallskip

\noindent Our results show that for the small levels of significance (such as 5\% or 1\%) of the Hausman pretest commonly used in applications,
the minimum coverage probability of the confidence interval for the slope parameter with nominal coverage probability $1-\alpha$ can
be far below nominal. The methodology that we have described makes it easy to assess, for a wide variety of circumstances, the effect of the
Hausman pretest on the minimum coverage probability of this confidence interval. An interesting finding is that if we increase the significance
level of the Hausman pretest to, say, 50\% then this minimum coverage probability is much closer to the nominal coverage $1-\alpha$
for a wide range of parameters. This suggests that the Hausman pretest might continue to be used in practice to good effect, provided
that one uses such a relatively high level of significance for this pretest.

\bigskip

%\newpage

\baselineskip=18pt

\begin{center}
\textbf{APPENDIX A. DEFINITION OF THE NON-EXOGENEITY PARAMETER $\boldsymbol{\tau}$}
\end{center}
%\medskip

%Suppose that $(\mu_i, x_{i1}, \dots, x_{iT})$ has a multivariate normal distribution with mean 0 and the the covariance matrix
%\eqref{CovMatrixMuiXit}.
\noindent For the compound symmetry case, it may be shown that the distribution of $\mu_i$ conditional on $(x_{i1}, \dots, x_{iT})$
is normal with mean
\begin{equation*}
\frac{\sigma_{\mu} \, \widetilde{\tau} \, T}{\big(1 + (T-1) \rho \big) \, \sigma_x} \, \overline{x}_i,
\end{equation*}
where $\overline{x}_i = T^{-1} \sum_{t=1}^T x_{it}$, and variance
\begin{equation*}
\sigma_{\mu}^2 \left(1 -\frac{\widetilde{\tau}^2 \, T}{1 + (T-1) \rho} \right).
\end{equation*}
This suggests that $\tau = \text{Corr}(\mu_i, \overline{x}_i)$ is a reasonable measure of the dependence between
$\mu_i$ and $(x_{i1}, \dots, x_{iT})$ i.e. that it is reasonable to designate $\tau$ as the non-exogeneity parameter.
It may be shown that $\text{Var}(\overline{x}_i) = \sigma_x^2 (1 + (T-1) \rho)/T$ and
$\text{Cov}(\mu_i, \overline{x}_i) = \widetilde{\tau} \, \sigma_{\mu} \, \sigma_x$.
Thus
\begin{equation*}
\tau = \widetilde{\tau} \left( \frac{T}{1+(T-1)\rho} \right)^{1/2}.
\end{equation*}
%
%and that
%
%\begin{equation*}
%\begin{bmatrix}
%\mu_i \\
%\overline{x}_i
%\end{bmatrix}
%\sim N
%\Bigg(
%0,
%\begin{bmatrix}
%\sigma^2_{\mu} & \tau \sigma_{\mu} \sigma_{\overline{x}} \\
%\tau \sigma_{\mu} \sigma_{\overline{x}} & \sigma^2_{\overline{x}}
%\end{bmatrix}
%\Bigg).
%\end{equation*}
%

For the first order autoregression case, it may be shown that the distribution of $\mu_i$ conditional on $(x_{i1}, \dots, x_{iT})$
is normal with mean
\begin{equation*}
\frac{\sigma_{\mu} \, \widetilde{\tau} \, T \, (1-\rho)}{(1 + \rho ) \, \sigma_x} \, \overline{\overline{x}}_i,
\end{equation*}
where $\overline{\overline{x}}_i = T^{-1} \, \left( (1-\rho)^{-1}(x_{i1} + x_{iT}) + \sum_{t=2}^{T-1}x_{it} \right)$, and variance
\begin{equation*}
\sigma_{\mu}^2 \left(1 -\frac{\widetilde{\tau}^2 \, (2 + (T-2) (1-\rho))}{1 + \rho} \right).
\end{equation*}
This suggests that $\tau = \text{Corr}(\mu_i, \overline{\overline{x}}_i)$ is a reasonable measure of the dependence between
$\mu_i$ and $(x_{i1}, \dots, x_{iT})$ i.e. that it is reasonable to designate $\tau$ as the non-exogeneity parameter.
It may be shown that
\begin{equation*}
\text{Var}(\overline{\overline{x}}_i) = \frac{ \sigma^2_x}{T^2} \, \frac{(T(1-\rho)
+ 2 \rho )(1+\rho)}{(1-\rho)^2}
\end{equation*}
and
\begin{equation*}
\text{Cov}(\mu_i, \overline{\overline{x}}_i)
= \widetilde{\tau} \, \sigma_{\mu} \, \sigma_{x} \, \frac{T (1 - \rho) + 2 \rho}{T \, |1 - \rho|}.
\end{equation*}
Thus
\begin{equation*}
\tau =\widetilde{\tau} \left( \frac{ T(1-\rho)+ 2\rho}{1+\rho} \right)^{1/2}.
\end{equation*}
%
%and that
%
%\begin{equation*}
%\begin{bmatrix}
%\mu_i \\
%\overline{\overline{x}}_i
%\end{bmatrix}
%\sim N
%\Bigg(
%0,
%\begin{bmatrix}
%\sigma^2_{\mu} & \tau \sigma_{\mu} \sigma_{\overline{\overline{x}}} \\
%\tau \sigma_{\mu} \sigma_{\overline{\overline{x}}} & \sigma^2_{\overline{\overline{x}}}
%\end{bmatrix}
%\Bigg).
%\end{equation*}
%

%\newpage

%\bigskip

\begin{center}
\textbf{APPENDIX B. DESCRIPTION OF THE ESTIMATORS OF THE RANDOM ERROR AND RANDOM EFFECT VARIANCES CONSIDERED}
\end{center}
%\medskip

\noindent It has been suggested in the literature (see e.g. Hsiao, 1986 and Baltagi, 2005) that if a negative estimate of variance is observed then
one should do as Maddala and Mount (1973) suggest and replace this negative estimate by 0. We use this kind of approach to ensure
that $\widehat{\sigma}^2_{\varepsilon}$  is always positive and $\widehat{\sigma}^2_{\mu}$ is always nonnegative. This ensures that
the proofs of Theorems 1 and 2 carry through for each of the three pairs of estimators that we consider in this paper. We
consider the following pairs of estimators of $\sigma_{\varepsilon}^2$ and $\sigma_{\mu}^2$:

\medskip

\noindent (1) \ The usual unbiased estimators. Define
\begin{equation*}
\widehat{\sigma}_{\varepsilon}^2 = \frac{1}{N(T-1)-1} \sum_{i=1}^N \sum_{t=1}^T r_{it}^2
\end{equation*}
and
$
\widehat{\sigma}^2_{\mu} = \max(0, \, \widetilde{\sigma}^2_{\mu}),
$
where
\begin{equation*}
\widetilde{\sigma}^2_{\mu} = \frac{1}{N-2} \sum_{i=1}^N \overline{r}_i^2 - \frac{1}{NT(T-1)-T} \sum_{i=1}^N \sum_{t=1}^T r_{it}^2.
\end{equation*}
The $r_{it}$'s are the OLS residuals from model \eqref{consistent_est_beta_model} and the $\overline{r}_i$'s are the OLS residuals from model \eqref{average_over_t_model}.
Note that $\widetilde{\sigma}^2_{\mu}$ is an unbiased estimator of $\sigma^2_{\mu}$ only for $\tau=0$.

\medskip

\noindent (2) \ Hsiao's (1986) maximum likelihood estimators $\widehat{\sigma}^2_{\varepsilon}$ and
$\sigma^2_{\mu}$. We assume, of course, that the maximum likelihood estimator is obtained by maximizing the log-likelihood
function subject to the parameter constraints $\sigma_{\varepsilon}^2 \ge 0$ and $\sigma_{\mu}^2 \ge 0$.

 \medskip

\noindent (3) \   Wooldridge's (2002) estimators. Define
\begin{equation*}
\widehat{\sigma}^2_{\varepsilon} = \max( -\epsilon \, \widetilde{\sigma}^2_{\varepsilon}, \, \widetilde{\sigma}^2_{\varepsilon})
\end{equation*}
where $\epsilon$ is a very small positive number and
\begin{equation*}
\widetilde{\sigma}_{\varepsilon}^{2} = \frac{1}{NT-K} \sum_{i=1}^N \sum_{t=1}^T \tilde{r}_{it}^2 - \frac{1}{NT(T-1)/2 - K} \sum_{i=1}^N \sum_{t=1}^{T-1} \sum_{s=t + 1}^T \tilde{r}_{it} \tilde{r}_{is} \, .
\end{equation*}
Also define
$
\widehat{\sigma}^2_{\mu} = \max(0, \, \widetilde{\sigma}^2_{\mu})
$
where
\begin{equation*}
\widetilde{\sigma}_{\mu}^2 = \frac{1}{NT(T-1)/2 -K} \sum_{i=1}^N \sum_{t=1}^{T-1} \sum_{s=t + 1}^T \tilde{r}_{it} \tilde{r}_{is} \, .
\end{equation*}
Here, the $\tilde{r}_{it}$'s are the residuals from pooled OLS estimation for the model \eqref{model}
and $K = 0$ (no d.o.f. correction) or $K = 2$ (d.o.f. correction).

%\newpage

%\bigskip

\begin{center}
\textbf{APPENDIX C. PROOFS OF THEOREMS 1, 2 AND 3}
\end{center}
%\medskip

\noindent The proofs in this section make use of the Hausman test statistic $H(\widehat{\sigma}_{\varepsilon},\widehat{\sigma}_{\mu})$ considered in this paper and the unbiased estimators of $\sigma_{\varepsilon}$ and $\sigma_{\mu}$
described in Appendix B.  It is important to note that there are three different test statistics that can be used to carry out the Hausman test in the panel data context.  Theorems 1, 2 and 3 hold for the three Hausman test statistics given by Hausman and Taylor (1981) and the three pairs of estimators described in Appendix B.  Proofs of these theorems using these other test statistics or estimators are omitted for the sake of brevity, but follow similar arguments to the proofs that we present.

%\bigskip
\medskip

\noindent \textbf{Proof of Theorem \ref{thm: CovProb_DependOnPsi}}

\smallskip

%\medskip

We present the proof of this result for the case that $\sigma_{\varepsilon}$ and $\sigma_{\mu}$ are replaced by the unbiased estimators
described in Appendix B.
Suppose that $N$, $T$, the level of significance $\widetilde{\alpha}$ of the Hausman pretest, the nominal coverage $1-\alpha$, $x$, $\sigma_{\varepsilon}$ and $\sigma_{\mu}$ are given.  Let
$\varepsilon = (\varepsilon_{i1}, \dots, \varepsilon_{iT}, \dots, \varepsilon_{N1}, \dots, \varepsilon_{NT})$ and
$\mu = (\mu_1, \dots, \mu_N)$.
Recall the random variables defined in Section 3,
$g_I = \big(\widehat{\beta}(\psi)-\beta \big)/\big(\text{Var}_0(\widehat{\beta}(\psi)|x) \big)^{1/2}$,
$g_J = (\widetilde{\beta}_W-\beta)/\big(\text{Var}(\widetilde{\beta}_W|x)\big)^{1/2}$ and
$h = (\widetilde{\beta}_W - \widetilde{\beta}_B)/\big(\text{Var}(\widetilde{\beta}_W|x) + \text{Var}_0(\widetilde{\beta}_B|x)\big)^{1/2}$.
Note that $g_I$, $g_J$ and $h$ are determined by $\sigma_{\varepsilon}$, $\sigma_{\mu}$ and $(x, \varepsilon, \mu)$.  Let $\widehat{g}_I$, $\widehat{g}_J$ and $\widehat{h}$ denote the statistics $g_I$, $g_J$ and $h$ when $\sigma_{\varepsilon}$ and $\sigma_{\mu}$ are replaced by the
unbiased estimators $\widehat{\sigma}_{\varepsilon}$ and $\widehat{\sigma}_{\mu}$ described in Appendix B.  We express $\widehat{g}_I$, $\widehat{g}_J$ and $\widehat{h}$ in terms of $(x, \varepsilon, \mu)$ and we emphasize this by using the notation $\widehat{g}_I(x, \varepsilon, \mu)$, $\widehat{g}_J(x, \varepsilon, \mu)$ and $\widehat{h}(x, \varepsilon, \mu)$, respectively.

Since $\widetilde{\beta}_B$ is defined to be the OLS estimator of $\beta$ based on the model \eqref{average_over_t_model},
\begin{align}
\label{expr_beta_tilde_B_simple}
\widetilde{\beta}_B
= \frac{\sum_{i=1}^N(\overline{x}_i - \overline{x})(\overline{y}_i - \overline{y})}{\sum_{i=1}^N (\overline{x}_i - \overline{x})^2}
= \beta + \frac{\sum_{i=1}^N (x_i - \overline{x}) \big( (\mu_i - \overline{\mu}) + (\overline{\varepsilon}_i - \overline{\varepsilon}) \big)}
{\SSB}
\end{align}
where $\overline{x} = (NT)^{-1} \sum_{i=1}^N \sum_{t=1}^T x_{it}$, $\overline{y} = (NT)^{-1} \sum_{i=1}^N \sum_{t=1}^T y_{it}$,
$\SSB = \sum_{i=1}^N(\overline{x}_i - \overline{x})^2$, $\overline{\mu} = N^{-1} \sum_{i=1}^N \mu_i$
and $\overline{\varepsilon} = (NT)^{-1} \sum_{i=1}^N \sum_{t=1}^T \varepsilon_{it}$.
Also, since $\widetilde{\beta}_W$ is the OLS estimator of $\beta$ based on the model \eqref{consistent_est_beta_model},
\begin{align}
\label{expr_beta_W_simple}
\widetilde{\beta}_W
= \frac{\sum_{i=1}^N \sum_{t=1}^T (x_{it} - \overline{x}_i)(y_{it} - \overline{y}_i)}{\sum_{i=1}^N \sum_{t=1}^T (x_{it} - \overline{x}_i)^2}
= \beta + \frac{\sum_{i=1}^N \sum_{t=1}^T (x_{it} - \overline{x}_i)(\varepsilon_{it} - \overline{\varepsilon}_i)}
{\SSW}
\end{align}
where $\SSW = \sum_{i=1}^N \sum_{t=1}^T (x_{it} - \overline{x}_i)^2$.
In our context, Maddala's (1971) equation (1.3) is
$\widehat{\beta}(\psi) = w(\psi) \widetilde{\beta}_W + (1-w(\psi))\widetilde{\beta}_B$,
where $w(\psi) = 1/(1+r(x)/q(\psi, T))$.
Thus $\widehat{\beta}(\widehat{\psi}) = \widehat{w}(\widehat{\psi}) \widetilde{\beta}_W + (1-\widehat{w}(\widehat{\psi}))\widetilde{\beta}_B$,
where $\widehat{\psi}= \widehat{\sigma}_{\mu} / \widehat{\sigma}_{\varepsilon}$.  Observe that $w(\widehat{\psi})/(1+w(\widehat{\psi})) = q(\widehat{\psi}, T)/r(x)$, where $r(x) = \SSB/\SSW$.
It follows from this and \eqref{expr_beta_tilde_B_simple} and \eqref{expr_beta_W_simple} that
\begin{align*}
\widehat{g}_I(x, \varepsilon, \mu) &=
 \frac{\displaystyle{\sum_{i=1}^N (\overline{x}_i - \overline{x})(\mu_i - \overline{\mu}_i + \overline{\varepsilon}_i - \overline{\varepsilon}) + q(\widehat{\psi}, T)\sum_{i=1}^N \sum_{t=1}^T (x_{it} - \overline{x}_i)(\varepsilon_{it} - \overline{\varepsilon}_i)}}
{\widehat{\sigma}_{\varepsilon} \, \left(q(\widehat{\psi},T) \, \SSW \, \left(r(x)+q(\widehat{\psi},T)\right)\right)^{1/2}}
\end{align*}
\begin{align*}
\widehat{g}_J(x, \varepsilon, \mu)
&= \frac{\displaystyle{\sum_{i=1}^N \sum_{t=1}^T (x_{it} - \overline{x}_i)(\varepsilon_{it} - \overline{\varepsilon}_i)}}
{\widehat{\sigma}_\varepsilon \left(\SSW\right)^{1/2}} \\
\widehat{h}(x, \varepsilon, \mu)
&= \frac{\displaystyle{ r(x) \sum_{i=1}^N \sum_{t=1}^T (x_{it} - \overline{x}_i)(\varepsilon_{it} - \overline{\varepsilon}_i) - \sum_{i=1}^N(\overline{x}_i - \overline{x})(\mu_i - \overline{\mu} + \overline{\varepsilon}_i - \overline{\varepsilon})}}
{ \widehat{\sigma}_{\varepsilon} \left( \SSB\left(r(x) + q(\widehat{\psi}, T)\right) \right)^{1/2}}.
\end{align*}
%
%where $\overline{x} = (NT)^{-1} \sum_{i=1}^N \sum_{t=1}^T x_{it}$, $\overline{\mu} = N^{-1} \sum_{i=1}^N  \mu_i$,
%$\overline{\varepsilon} = (NT)^{-1} \sum_{i=1}^N \sum_{t=1}^T \varepsilon_{it}$.

% Introduce all notation for proof here.
We now introduce the following notation.  Let $x_{it}^{\dag} = x_{it} / \sigma_x$, $\overline{x}_i^{\dag} = T^{-1} \sum_{t=1}^T x_{it}^{\dag}$, $\overline{x}^{\dagger} = (NT)^{-1} \sum_{i=1}^N \sum_{t=1}^T x^{\dag}_{it}$, $\SSW^{\dag} = \sum_{i=1}^N \sum_{t=1}^T (x_{it}^{\dag} - \overline{x}_i^{\dag})^2$, $\SSB^{\dagger} = \sum_{i=1}^N(\overline{x}^{\dagger}_i - \overline{x}^{\dagger})^2$, $\varepsilon_{jt}^{\dag} = \varepsilon_{jt} / \sigma_{\varepsilon}$, $\overline{\varepsilon}_j^{\dag} = T^{-1} \sum_{t=1}^T \varepsilon_{jt}^{\dag}$,  $\overline{\varepsilon}^{\dagger} = (NT)^{-1} \sum_{i=1}^N \sum_{t=1}^T \varepsilon_{it}^{\dagger}$, $\mu_i^{\dagger} = \mu_i/\sigma_{\mu}$, $\overline{\mu}^{\dagger}= N^{-1}\sum_{i=1}^N \mu_i^{\dagger}$ and $x^{\dag} = (x_{11}^{\dag}, \dots, x_{1T}^{\dag}, x_{21}^{\dag}, \dots, x_{2T}^{\dag}, \dots, x_{N1}^{\dag}, \dots, x_{NT}^{\dag})$.  Thus $r(x) = \SSB/\SSW = \SSB^{\dag}/\SSW^{\dag} =: r(x^{\dag}) $ and $q(\widehat{\psi}, T) = \widehat{\psi}^2 + (1/T)$.

Note that the $\varepsilon_{it}^{\dag}$'s and the $(\mu_i^{\dag}, x_{i1}^{\dag}, \dots, x_{iT}^{\dag})$'s are independent, the
$(\mu_i^{\dag}, x_{i1}^{\dag}, \dots, x_{iT}^{\dag})$'s are
i.i.d., the $\varepsilon_{it}^{\dag}$'s are i.i.d. $N(0, 1)$
and the $\mu_i^{\dag}$'s are i.i.d. $N(0, 1)$. Also note that
the distribution of $(\mu_i^{\dag}, x_{i1}^{\dag}, \dots, x_{iT}^{\dag})$ is
multivariate normal with mean $0$ and covariance matrix
\begin{equation*}
\left[\begin{matrix} 1 \quad \widetilde{\tau} e' \\
\widetilde{\tau} e \quad G \end{matrix} \right],
\end{equation*}
where $e$ is a $T$-vector of 1's, and $\widetilde{\tau}$ is described in Appendix A for both the compound symmetry
and the first order autoregression cases.  Thus the joint distribution of the
$\varepsilon_{it}^{\dag}$'s and the $(\mu_i^{\dag}, x_{i1}^{\dag}, \dots, x_{iT}^{\dag})$'s
is determined by $\rho$ and $\tau$ (and does not depend on either $\sigma_{\varepsilon}$ or $\sigma_{\mu}$ or $\sigma_x$).

We now show that $\widehat{g}_I(x, \varepsilon, \mu)$, $\widehat{g}_J(x, \varepsilon, \mu)$ and $\widehat{h}(x, \varepsilon, \mu)$ can be written in terms of the $x_{it}^{\dag}$'s, $\varepsilon_{it}^{\dag}$'s, $\mu_i^{\dag}$'s and $\psi$. This has the consequence that both the distribution of $(\widehat{g}_I(x, \varepsilon, \mu), \widehat{h}(x, \varepsilon, \mu))$ and the distribution of $(\widehat{g}_J(x, \varepsilon, \mu), \widehat{h}(x, \varepsilon, \mu))$ are functions of the quantities
$N$, $T$, $\widetilde{\alpha}$ and $1-\alpha$ and the unknown parameters $\psi$, $\rho$ and $\tau$. The theorem follows from this.

The $i$'th OLS residual at time $t$ from model \eqref{consistent_est_beta_model}, which appears in the
expressions for the
usual unbiased estimators of $\sigma^2_{\varepsilon}$ and $\sigma^2_{\mu}$ is
\begin{align*}
r_{it} = (\varepsilon_{it} - \overline{\varepsilon}_i) - (x_{it} - \overline{x}_i)
\frac{\sum_{j=1}^N \sum_{s=1}^T (x_{js} - \overline{x}_j)(\varepsilon_{js} - \overline{\varepsilon}_j)}
{\SSW}.
\end{align*}
Obviously,
\begin{align*}
r_{it} = (\varepsilon_{it} - \overline{\varepsilon}_i) - (x_{it}^{\dag} - \overline{x}_i^{\dag})
\frac{\sum_{j=1}^N \sum_{s=1}^T (x_{js}^{\dag} - \overline{x}_j^{\dag})(\varepsilon_{js} - \overline{\varepsilon}_j)}
{\SSW^{\dag}}.
\end{align*}
Dividing by $\sigma_{\varepsilon}$ gives
\begin{align*}
\frac{r_{it}}{\sigma_{\varepsilon}}
&= (\varepsilon_{it}^{\dag} - \overline{\varepsilon}_i^{\dag}) - (x_{it}^{\dag} - \overline{x}_i^{\dag})
\frac{\sum_{j=1}^N \sum_{s=1}^T (x_{js}^{\dag} - \overline{x}_j^{\dag})(\varepsilon_{js}^{\dag} - \overline{\varepsilon}_j^{\dag})}
{\SSW^{\dag}}.
\end{align*}
Hence
$\widehat{\sigma}_{\varepsilon}^2 / \sigma_{\varepsilon}^2$ is a function of the $x_{it}^{\dag}$'s and the
$\varepsilon_{it}^{\dag}$'s.  In a similar manner, it can be shown that $r_{it}/\sigma_{\mu}$ and $\overline{r}_i/\sigma_{\mu}$ are functions of the $x^{\dag}_{it}$'s, $\varepsilon^{\dag}_{it}$'s, $\mu_i^{\dag}$'s and $\psi$, where $\overline{r}_i$ is defined in Appendix B.  Hence $\widetilde{\sigma}_{\mu}^2/\sigma^2_{\mu}$ is a function of the $x^{\dag}_{it}$'s, $\varepsilon^{\dag}_{it}$'s, $\mu_i^{\dag}$'s and $\psi$.  Thus $\widehat{\sigma}^2_{\mu}/\sigma^2_{\mu}$ is also a function of the $x^{\dag}_{it}$'s, $\varepsilon^{\dag}_{it}$'s, $\mu_i^{\dag}$'s and $\psi$.  Now
\begin{equation*}
\widehat{\psi}
= \widehat{\sigma}_{\mu} / \widehat{\sigma}_{\varepsilon}
= \psi \, \frac{\widehat{\sigma}_{\mu} / \sigma_{\mu}}{\widehat{\sigma}_{\varepsilon} / \sigma_{\varepsilon}}.
\end{equation*}
Therefore,  $\widehat{\psi}$ is a function of
the $x^{\dag}_{it}$'s, $\varepsilon^{\dag}_{it}$'s, $\mu_i^{\dag}$'s and $\psi$.

Dividing the numerator and denominator of the expression for $\widehat{h}(x, \varepsilon, \mu)$
by $\sigma_{\varepsilon}$, we obtain
\begin{align*}
\widehat{h}(x, \varepsilon, \mu) = \frac{\displaystyle{ r(x^{\dag}) \sum_{i=1}^N \sum_{t=1}^T (x_{it} - \overline{x}_i)(\varepsilon^{\dag}_{it} - \overline{\varepsilon}^{\dag}_i) - \sum_{i=1}^N(\overline{x}_i - \overline{x})(\psi(\mu^{\dag}_i - \overline{\mu}^{\dag}) + \overline{\varepsilon}^{\dag}_i - \overline{\varepsilon}^{\dag})}}
{ (\widehat{\sigma}_{\varepsilon}/\sigma_{\varepsilon}) \left( \SSB\left(r(x^{\dag}) + q(\widehat{\psi}, T)\right) \right)^{1/2}}.
\end{align*}
Multiplying the numerator and denominator
by $\sigma_x$, we obtain
\begin{align*}
%\frac{\frac{\displaystyle{\sum_{i=1}^N \sum_{t=1}^T (x_{it}^{\dag} - \overline{x}_i^{\dag})(\varepsilon_{it}^{\dag} - \overline{\varepsilon}_i^{\dag})}}{\SSW^{\dag}}
%\, - \, \frac{\displaystyle{\sum_{i=1}^N(\overline{x}_i^{\dag} - \overline{x}^{\dag})
%\big(\psi(\mu_i^{\dag} - \overline{\mu}^{\dag}) + \overline{\varepsilon}_i^{\dag} - \overline{\varepsilon}^{\dag} \big)}}{\SSB^{\dag}}}
%{\displaystyle{\frac{\widehat{\sigma}_{\varepsilon}}{\sigma_{\varepsilon}}}
%\left(\displaystyle{\frac{1}{\SSW^{\dag}} + \frac{q(\widehat{\psi}, T)}{\SSB^{\dag}}}\right)^{1/2}}. \\
\widehat{h}(x, \varepsilon, \mu) = \frac{\displaystyle{ r(x^{\dag}) \sum_{i=1}^N \sum_{t=1}^T (x^{\dag}_{it} - \overline{x}^{\dag}_i)(\varepsilon^{\dag}_{it} - \overline{\varepsilon}^{\dag}_i) - \sum_{i=1}^N(\overline{x}^{\dag}_i - \overline{x}^{\dag})(\psi(\mu^{\dag}_i - \overline{\mu}^{\dag}) + \overline{\varepsilon}^{\dag}_i - \overline{\varepsilon}^{\dag})}}
{ (\widehat{\sigma}_{\varepsilon}/\sigma_{\varepsilon}) \left( \SSB^{\dag}\left(r(x^{\dag}) + q(\widehat{\psi}, T)\right) \right)^{1/2}}.
\end{align*}
Therefore, $\widehat{h}(x, \varepsilon, \mu)$ is a function of
the $x^{\dag}_{it}$'s, $\varepsilon^{\dag}_{it}$'s, $\mu_i^{\dag}$'s and $\psi$.

In a similar manner, it can be shown that
\begin{equation*}
\widehat{g}_I(x, \varepsilon, \mu)
= \frac{\displaystyle{\sum_{i=1}^N (\overline{x}_i^{\dag} - \overline{x}^{\dag})(\psi(\mu_i^{\dag} - \overline{\mu}_i^{\dag}) + \overline{\varepsilon}_i^{\dag} - \overline{\varepsilon}^{\dag}) + q(\widehat{\psi}, T)\sum_{i=1}^N \sum_{t=1}^T (x_{it}^{\dag} - \overline{x}_i^{\dag})(\varepsilon_{it}^{\dag} - \overline{\varepsilon}_i^{\dag})}}
{\widehat{\psi} \, \big(q(\widehat{\psi},T) \, \SSW^{\dag} \, (r(x^{\dag})+q(\widehat{\psi},T))\big)^{1/2}}.
\end{equation*}
and that
\begin{equation*}
\widehat{g}_J(x, \varepsilon, \mu)
= \frac{\displaystyle{\sum_{i=1}^N \sum_{t=1}^T (x^{\dag}_{it} - \overline{x}^{\dag}_i)(\varepsilon^{\dag}_{it} - \overline{\varepsilon}^{\dag}_i)}}
{\displaystyle{(\widehat{\sigma}_\varepsilon/\sigma_{\varepsilon})} \left(\SSW^{\dag}\right)^{1/2}}.
\end{equation*}
Thus $\widehat{g}_I(x, \varepsilon, \mu)$ and $\widehat{g}_J(x, \varepsilon, \mu)$ are also functions of
the $x^{\dag}_{it}$'s, $\varepsilon^{\dag}_{it}$'s, $\mu_i^{\dag}$'s and $\psi$.

\bigskip

\noindent \textbf{Proof of Theorem \ref{thm: eveness}}

\medskip

Suppose that $N$, $T$, the nominal level of significance $\widetilde{\alpha}$ of the Hausman pretest, the nominal coverage $1-\alpha$, $x$,
$\sigma_{\varepsilon}$ and $\sigma_{\mu}$  are fixed.
We assume that $(\mu_i, x_{i1}, \dots, x_{iT})$ has a multivariate normal distribution
with mean $0$ and the covariance matrix \eqref{CovMatrixMuiXit} where $\widetilde{\tau} = \tau/\sqrt{T}$ and $G = I$ (the $T\times T$ identity matrix).
The proof when $G$ has either a compound symmetry or first order autoregression structure and $\rho \ne 0$
follows in a similar manner using the definitions of $\tau$ given in Appendix A.

By Theorem 1 the coverage probability of the confidence interval constructed after a Hausman pretest is a function of $\tau$.
So, in this section $P_{\tau = d}(A \, | \, x)$ denotes the probability of the event $A$ conditional on $x$, evaluated at $\tau = d$. The set of possible values of $\tau$ is $(-1, 1)$.
Our aim is to show that  the coverage probability is an even function of $\tau$, i.e. our aim is to show that
\begin{equation*}
%\label{CovProbEvenFunction}
P_{\tau = d}(\beta \in K(\widehat{\sigma}_{\varepsilon}, \widehat{\sigma}_{\mu}) ) = P_{\tau = -d}(\beta \in K(\widehat{\sigma}_{\varepsilon}, \widehat{\sigma}_{\mu}) ).
\end{equation*}
for every $\tau \in (0,1)$.
By the law of total probability, $P \big(\beta \in K(\widehat{\sigma}_{\varepsilon},\widehat{\sigma}_{\mu}) \big)$ is equal to
\begin{equation*}
 P \big(\beta \in I(\widehat{\sigma}_{\varepsilon},\widehat{\sigma}_{\mu}), \,  H(\widehat{\sigma}_{\varepsilon}, \widehat{\sigma}_{\mu}) \leq z^2_{1-\widetilde{\alpha}/2}\big)
+ P \big(\beta \in J(\widehat{\sigma}_{\varepsilon}), \, H(\widehat{\sigma}_{\varepsilon},\widehat{\sigma}_{\mu}) > z^2_{1-\widetilde{\alpha}/2}  \big).
\end{equation*}
Using the definitions of $\widehat{g}_I$, $\widehat{g}_J$ and $\widehat{h}$ stated in the proof of Theorem 1,
$P \big(\beta \in I(\widehat{\sigma}_{\varepsilon}, \widehat{\sigma}_{\mu}),
\, H(\widehat{\sigma}_{\varepsilon}, \widehat{\sigma}_{\mu}) \leq z^2_{1-\widetilde{\alpha}/2}\big)$ is equal to
$P \big( |\widehat{g}_I| \le z_{1-\alpha/2}, \, |\widehat{h}| \le z_{1-\widetilde{\alpha}/2} \big)$
and by the law of total probability, $P \big(\beta \in J(\widehat{\sigma}_{\varepsilon}), \, H(\widehat{\sigma}_{\varepsilon},\widehat{\sigma}_{\mu}) > z^2_{1-\widetilde{\alpha}/2}  \big)$ is equal to
$P\big(|\widehat{g}_J| \leq z_{1-\alpha/2}\big) - P\big( |\widehat{g}_J| \leq z_{1-\alpha/2},\,  |\widehat{h}| \leq z_{1-\widetilde{\alpha}/2}\big)$.
Therefore, to prove that the coverage probability is an even function of $\tau$, when $\sigma_{\varepsilon}$ and $\sigma_{\mu}$ are unknown, it is sufficient to prove the following:
\begin{align*}
\text{(a)} & \, \,  P_{\tau = d} \big( |\widehat{g}_I| \le z_{1-\alpha/2}, \,
 |\widehat{h}| \le z_{1-\widetilde{\alpha}/2} \big)
 = P_{\tau = -d} \big( |\widehat{g}_I| \le z_{1-\alpha/2}, \,
 |\widehat{h}| \le z_{1-\widetilde{\alpha}/2} \big), \\
\text{(b)} & \, \, P_{\tau = d}\big( |\widehat{g}_J| \leq z_{1-\alpha/2},\, |\widehat{h}| \leq z_{1-\widetilde{\alpha}/2}\big)
 =P_{\tau = -d}\big(|\widehat{g}_J| \leq z_{1-\alpha/2}, \, |\widehat{h}| \leq z_{1-\widetilde{\alpha}/2}\big) \text{ and} \\
\text{(c)} & \, \,
P_{\tau = d}\big( |\widehat{g}_J| \leq z_{1-\alpha/2}) = P_{\tau = -d}( |\widehat{g}_J| \leq z_{1-\alpha/2}\big).
\end{align*}

First we consider (a).  We have that
\begin{align*}
\{ -z_{1-\alpha/2} \leq \widehat{g}_I \leq z_{1-\alpha/2} \} = \{ z_{1-\alpha/2} \geq -\widehat{g}_I \geq -z_{1-\alpha/2} \}
= \{ -z_{1-\alpha/2} \leq -\widehat{g}_I \leq z_{1-\alpha/2} \}.
\end{align*}
Similarly,
$
\{ -z_{1-\widetilde{\alpha}/2} \leq \widehat{h} \leq z_{1-\widetilde{\alpha}/2} \} = \{ -z_{1-\widetilde{\alpha}/2} \leq -\widehat{h} \leq z_{1-\widetilde{\alpha}/2} \}.
$
Therefore, to show that (a) holds, it is sufficient to show that
\begin{align*}
&P_{\tau = d} \big( - z_{1-\alpha/2}  \le  \widehat{g}_I \le z_{1-\alpha/2}, \,
-z_{1-\widetilde{\alpha}/2} \le \widehat{h} \le z_{1-\widetilde{\alpha}/2} \big) \nonumber \\
&= P_{\tau = -d} \big( - z_{1-\alpha/2}  \le  -\widehat{g}_I \le z_{1-\alpha/2}, \,
-z_{1-\widetilde{\alpha}/2} \le -\widehat{h} \le z_{1-\widetilde{\alpha}/2} \big) \label{eveness: a},
\end{align*}

We introduce the following notation.
Let  $x_{it}^* = -x_{it}$ for $ i = 1, \dots, N$, $t=1, \dots, T$ and $x^* = (-x_{11}, \dots, -x_{1T}, -x_{21}, \dots, -x_{2T}, -x_{N1}, \dots, -x_{NT})$.  Let $\overline{x}_i^* = T^{-1}\sum_{t=1}^Tx_{it}^*$, $\overline{x}^* = (NT)^{-1}\sum_{i=1}^N \sum_{t=1}^T x_{it}^*$, $\SSB^* = \sum_{i=1}^N (\overline{x}_i^* - \overline{x}^*)^2$, $\SSW^* = \sum_{i=1}^N \sum_{t=1}^T (x_{it}^* - \overline{x}_i^*)^2$ and $r(x^*) = \SSB^*/\SSW^*$.  Note that $\SSB^* = \SSB$, $\SSW^* = \SSW$ and $r(x^*) = r(x)$.

For $\tau = d$, $(\mu_i, x_{i1}, \dots, x_{iT})$ has a multivariate normal distribution
with mean $0$ and covariance matrix \eqref{CovMatrixMuiXit},
where $\widetilde{\tau} = d/\sqrt{T}$
and $G=I$.
Observe that,
for $\tau = d$, $(\mu_i, x_{i1}^*, \dots, x_{iT}^*)$ has a multivariate normal distribution
with mean $0$ and covariance matrix \eqref{CovMatrixMuiXit},
where $\widetilde{\tau} = -d/\sqrt{T}$
and $G=I$. Hence, for $\tau = -d$, $(\mu_i, x_{i1}^*, \dots, x_{iT}^*)$ has a multivariate normal distribution
with mean $0$ and covariance matrix \eqref{CovMatrixMuiXit},
where $\widetilde{\tau} = d/\sqrt{T}$ and $G=I$.

From this point onwards we write $\widehat{g}_I$ as $\widehat{g}_I(x, \varepsilon, \mu)$ and $\widehat{h}$ as $\widehat{h}(x, \varepsilon, \mu)$ (as in the proof of Theorem 1) to emphasize the dependence on $(x,\varepsilon,\mu)$.  Recall, from the proof of Theorem 1, that
\begin{equation*}
\widehat{g}_I(x, \varepsilon, \mu) =
 \frac{\displaystyle{\sum_{i=1}^N (\overline{x}_i - \overline{x})(\mu_i - \overline{\mu}_i + \overline{\varepsilon}_i - \overline{\varepsilon}) + q(\widehat{\psi}, T)\sum_{i=1}^N \sum_{t=1}^T (x_{it} - \overline{x}_i)(\varepsilon_{it} - \overline{\varepsilon}_i)}}
{\widehat{\sigma}_{\varepsilon} \, \big(q(\widehat{\psi},T) \, \SSW \, (r(x)+q(\widehat{\psi},T))\big)^{1/2}}.
\end{equation*}
It follows that
\begin{align*}
-\widehat{g}_I(x, \varepsilon, \mu) =
 \frac{\displaystyle{\sum_{i=1}^N (\overline{x}_i^* - \overline{x}^*)(\mu_i - \overline{\mu}_i + \overline{\varepsilon}_i - \overline{\varepsilon}) + q(\widehat{\psi}, T)\sum_{i=1}^N \sum_{t=1}^T (x^*_{it} - \overline{x}^*_i)(\varepsilon_{it} - \overline{\varepsilon}_i)}}
{\widehat{\sigma}_{\varepsilon} \, \big(q(\widehat{\psi},T) \, \SSW^* \, (r(x^*)+q(\widehat{\psi},T))\big)^{1/2}}.
\end{align*}

The usual unbiased estimator of $\sigma_{\varepsilon}^2$ is
\begin{equation*}
\widehat{\sigma}_{\varepsilon}^2
= \frac{1}{N(T-1)-1} \sum_{i=1}^N  \sum_{t=1}^T \big((\varepsilon_{it} - \overline{\varepsilon}_i) - (\widetilde{\beta}_W - \beta) (x_{it} - \overline{x}_i) \big)^2.
\end{equation*}
Note that
\begin{equation*}
\widetilde{\beta}_W - \beta = \frac{\sum_{j=1}^N \sum_{s=1}^T (x_{js} - \overline{x}_j)(\varepsilon_{js}-\overline{\varepsilon}_j)}{\SSW}.
\end{equation*}
Thus
\begin{align*}
(\widehat{\sigma}^*_{\varepsilon})^2 &= \frac{1}{N(T-1)-1} \sum_{i=1}^N \sum_{t=1}^T \left( (\varepsilon_{it} - \overline{\varepsilon}_i) - \frac{\sum_{j=1}^N \sum_{s=1}^T(x_{js}^* - \overline{x}_j^*)(\varepsilon_{js}-\overline{\varepsilon}_j)}{\SSW^*} (x_{it}^*-\overline{x}_i^*)\right)^2 \\
&= \widehat{\sigma}_{\varepsilon}^2
\end{align*}
We define $(\widetilde{\sigma}^*_{\mu})^2$ to be $\widetilde{\sigma}^2_{\mu}$, but with $x_{it}$
replaced by $x_{it}^*$ for $i=1, \dots, N$ and $t=1, \dots, T$. It can be shown that $(\widetilde{\sigma}^*_{\mu})^2 = \widetilde{\sigma}^2_{\mu}$.  Hence for the usual unbiased (for $\tau=0$) estimator $\widetilde{\sigma}^2_{\mu}$, $(\widehat{\sigma}^*_{\mu})^2 = \widehat{\sigma}_{\mu}^2$.
We then define $ \widehat{\psi}^* = \widehat{\sigma}_{\mu}^*/\widehat{\sigma}_{\varepsilon}^* $ and note that $ \widehat{\psi}^* = \widehat{\psi} $.

Hence
\begin{align*}
-\widehat{g}_I(x, \varepsilon, \mu) &=
\frac{\displaystyle{\sum_{i=1}^N (\overline{x}_i^* - \overline{x}^*)(\mu_i - \overline{\mu}_i + \overline{\varepsilon}_i - \overline{\varepsilon}) + q(\widehat{\psi}^*, T)\sum_{i=1}^N \sum_{t=1}^T (x^*_{it} - \overline{x}^*_i)(\varepsilon_{it} - \overline{\varepsilon}_i)}}
{\widehat{\sigma}_{\varepsilon}^* \, \big(q(\widehat{\psi}^*,T) \, \SSW^* \, (r(x^*)+q(\widehat{\psi}^*,T))\big)^{1/2}} \\
&= \widehat{g}_I(x^*, \mu, \varepsilon).
\end{align*}
By a similar argument,
\begin{align*}
\widehat{h}(x, \varepsilon, \mu)
&= \frac{\displaystyle{ r(x^*) \sum_{i=1}^N \sum_{t=1}^T (x^*_{it} - \overline{x}^*_i)(\varepsilon_{it} - \overline{\varepsilon}_i) - \sum_{i=1}^N(\overline{x}^*_i - \overline{x}^*)(\mu_i - \overline{\mu} + \overline{\varepsilon}_i - \overline{\varepsilon})}}
{ \widehat{\sigma}_{\varepsilon} \left( \SSB^*\left(r(x^*) + q(\widehat{\psi}^*, T)\right) \right)^{1/2}} \\
&= \widehat{h}(x^*, \varepsilon, \mu).
\end{align*}

We see that $-\widehat{g}_I(x, \varepsilon, \mu)$ and $-\widehat{h}(x, \varepsilon, \mu)$ are the same functions of the $x_{it}^*$'s, $\varepsilon_{it}^*$'s and $\mu_i^*$'s as $\widehat{g}_I(x, \varepsilon, \mu)$ and $\widehat{h}(x, \varepsilon, \mu)$, respectively, are functions of the $x_{it}$'s, $\varepsilon_{it}$'s and $\mu_i$'s.
Hence (a) is true.  In a similar manner, it can be shown that (b) and (c) are also true.  Therefore the coverage probability is an even function of $\tau$.

\bigskip

\noindent \textbf{Proof of Theorem 3}

\medskip

Suppose that the matrix $G$ has 1's on the diagonal and $\rho$ elsewhere (compound symmetry).
As shown in Appendix A, $\tau = \text{Corr}(\mu_i, \overline{x}_i)$ and so
\begin{equation*}
\begin{bmatrix}
\mu_i \\
\overline{x}_i
\end{bmatrix}
\sim N
\Bigg(
0,
\begin{bmatrix}
\sigma^2_{\mu} & \tau \sigma_{\mu} \sigma_{\overline{x}} \\
\tau \sigma_{\mu} \sigma_{\overline{x}} & \sigma^2_{\overline{x}}
\end{bmatrix}
\Bigg),
\end{equation*}
where
 $\sigma_{\overline{x}}^2 = \text{Var}(\overline{x}_i)$ (a convenient formula for $\text{Var}(\overline{x}_i)$ is given in Appendix A).
Therefore, conditional on $x$, $\mu_i = \tau (\sigma_{\mu}/\sigma_{\overline{x}}) \, \overline{x}_i  + \eta_i$,
%
%\begin{equation*}
%\label{conditional_model_mu_i1}
%\mu_i = \tau \frac{\sigma_{\mu}}{\sigma_{\overline{x}}} \, \overline{x}_i  + \eta_i,
%\end{equation*}
%
where the $\eta_i$'s and $\varepsilon_{it}$'s are independent and the $\eta_i$'s are i.i.d. $N \big(0, (1 - \tau^2) \sigma_{\mu}^2 \big)$.
It follows from this that, conditional on $x$,
\begin{equation}
\label{cond_mui_minus_mubar}
\mu_i - \overline{\mu}
= \tau \frac{\sigma_{\mu}}{\sigma_{\overline{x}}} \left( \overline{x}_i - \overline{x} \right ) + (\eta_i -\overline{\eta}).
\end{equation}

Consider the expression \eqref{expr_beta_tilde_B_simple} for $\widetilde{\beta}_B$. It follows from \eqref{cond_mui_minus_mubar}
that, conditional on $x$,
\begin{equation*}
\widetilde{\beta}_B
= \beta
+ \tau \frac{\sigma_{\mu}}{\sigma_{\overline{x}}}
+ \frac{\sum_{i=1}^N (\overline{x}_i - \overline{x})((\eta_i -\overline{\eta}) +  (\overline{\varepsilon}_i - \overline{\varepsilon}))}
{\SSB}.
\end{equation*}
Obviously, $E(\widetilde{\beta}_B \, | \, x) = \beta +\tau (\sigma_{\mu}/\sigma_{\overline{x}})$. It can be shown, after lengthy algebraic manipulations,
that
\begin{equation*}
\text{Var}(\widetilde{\beta}_B \, | \, x)
= \frac{\sigma^2_{\varepsilon}}{\SSW} \frac{(1-\tau^2)\psi^2 + (1/T)}{r(x)},
\end{equation*}
where $\SSW = \sum_{i=1}^N \sum_{t=1}^T (x_{it} - \overline{x}_i)^2$ and $r(x) = \SSB/\SSW$.

Now consider the expression \eqref{expr_beta_W_simple} for $\widetilde{\beta}_W$.
Obviously, $E(\widetilde{\beta}_W \, | \, x) = \beta$. It can be shown, after some algebraic manipulation, that
$\text{Var}(\widetilde{\beta}_W \, | \, x) = \sigma^2_{\varepsilon} / \SSW$. It can also be shown, after lengthy algebraic manipulations,
that $\text{Cov}(\widetilde{\beta}_B, \widetilde{\beta}_W \, | \, x) = 0$. We conclude that, conditional on $x$, $\widetilde{\beta}_B$ and
$\widetilde{\beta}_W$ are independent normally distributed random variables with the stated conditional means and variances.

The distributions of the random vectors $(g_I, h)$ and $(g_J, h)$ are determined by the bivariate normal distributions of $(\widetilde{\beta}_W - \beta, \, \widetilde{\beta}_W - \widetilde{\beta}_B)$ and $(\widehat{\beta}-\beta, \, \widetilde{\beta}_W - \widetilde{\beta}_B)$.  In our context, Maddala's (1971) equation (1.3) is $\widehat{\beta} = w(\psi) \widetilde{\beta}_W + (1-w(\psi))\widetilde{\beta}_B$, where $w(\psi) = 1/(1+r(x)/q(\psi, T))$.
It follows from this equation that the distributions of $(\widetilde{\beta}_W - \beta, \, \widetilde{\beta}_W - \widetilde{\beta}_B)$ and $(\widehat{\beta}-\beta, \, \widetilde{\beta}_W - \widetilde{\beta}_B)$, conditional on $x$, are bivariate normal, where
\begin{align*}
&E(\widetilde{\beta}_W - \beta \, | \, x) = 0, \, \, \,
\text{Var}(\widetilde{\beta}_W - \beta \, | \, x)
= \text{Cov}(\widetilde{\beta}_W - \beta, \widetilde{\beta}_W - \widetilde{\beta}_B \, | \, x) =\sigma^2_{\varepsilon}/\SSB, \\
&E(\widetilde{\beta}_W - \widetilde{\beta}_B | x)
= -\tau \frac{ \sigma_{\mu}}{\sigma_{\overline{x}}},
\, \, \, \text{Var}(\widetilde{\beta}_W - \widetilde{\beta}_B \, | \, x)
= \frac{\sigma^2_{\varepsilon}}{\SSW} \left( \frac{ (1-\tau^2)\psi^2 + (1/T)}{r(x)} + 1\right), \\
&E(\widehat{\beta} - \beta \, | \, x) = (1-w)\tau \frac{\sigma_{\mu}}{\sigma_{\varepsilon}}, \\
&\text{Var}(\widehat{\beta} - \beta \, | \, x)
= \frac{\sigma^2_{\varepsilon}}{\SSW} \left( \frac{ (1-w(\psi))^2(q(\psi, T) - \tau^2 \psi^2)}{r(x)} + w^2(\psi)\right) \\
& \text{and } \text{Cov}(\widehat{\beta}-\beta, \, \widetilde{\beta}_W - \widetilde{\beta}_B \, | \, x)
= \frac{\sigma^2_{\varepsilon}}{\SSW} \left( w(\psi) - \frac{(1-w(\psi))(1-\tau^2)\psi^2 + (1/T)}{r(x)}\right).
\end{align*}
Theorem 3 follows from these distributional properties.

\bigskip

%\newpage

\baselineskip=19pt

\begin{center}
\textbf{REFERENCES}
\end{center}

\medskip

\rf Ajmani, V.B., 2009. Applied Econometrics Using the SAS system. John Wiley, Hoboken, N.J.

\smallskip

\rf Baltagi, B.H., 2005. Econometric Analysis of Panel Data, 3rd edition. John Wiley \& Sons, Ltd.

\smallskip

\rf Bedard, K., Deschenes, O., 2006.  The long-term impact of military service on health: Evidence from work war II and Korean war veterans.  American Economic Review 96, 176-194.

\smallskip

\rf Bloningen, B.A., 1997. Firm-specific assets and the link between exchange rates and foreign direct investment.
American Economic Review 87, 447--465.

\smallskip

\rf Croissant, Y., Millo, G., 2008. Panel data econometrics in R: The plm package. Journal of Statistical Software, 27(2), 1-43.

\smallskip

\rf Ebbes, P., Bockenholt, U., Wedel, M., 2004. Regressor and random-effects dependencies in multilevel models. Statistica Neerlandica 58, 161--178.

\smallskip

\rf Freeman, P.R., 1989. The performance of the two-stage analysis of two-treatment, two-period crossover trials.
Statistics in Medicine 8, 1421--1432.

\smallskip

\rf Gardiner, J.C., Luo, Z., Roman, L.A., 2009.  Fixed effects, random effects and GEE: What are the differences? Statistics in Medicine 28, 221--239.

\smallskip

\rf Gaynor, M., Seider, H., Vogt, W.B., 2005.  The volume-outcome effect, scale economies, and learning-by-doing.  American Economic Review 95, 243-247.

\smallskip

\rf Griffiths, W. E., Hill, R. C., Lim, G., 2012. Using EViews for Principles of Econometrics. Danvers, MA: John Wiley \& Sons.

 \smallskip

\rf Guggenberger, P., 2010. The impact of a Hausman pretest on the size of a hypothesis test: The panel data case.
Journal of Econometrics 156, 337--343.

\smallskip

\rf Hastings, J.S., 2004. Vertical relationships and competition in retail gasoline markets: Empirical evidence from contract changes in Southern
California. American Economic Review 94, 317--328.

\smallskip

\rf Hsiao, C., 1986. Analysis of Panel Data. Cambridge University Press, Cambridge.

\smallskip

\rf Hausman, J.A., 1978. Specification tests in econometrics. Econometrica 46, 1251--1271.

\smallskip

\rf  Hausman, J.A., Taylor, W.E., 1981.  Panel data and unobservable individual effects.  Econometrica 49, 1377--1398.

\smallskip

\rf  Jackowicz, K., Kowalewski, O., Kozlowski, L., 2013.  The influence of political factors on commercial banks in Central European countries. Journal of Financial Stability 9, 759--777.

\smallskip

\rf Kabaila, P., 1995. The effect of model selection on confidence regions and prediction regions. Econometric Theory 11, 537--549.

\smallskip

\rf Kabaila, P., 2009. The coverage properties of confidence regions after model
selection. International Statistical Review 77, 405--414.

\smallskip

\rf Kabaila, P., Leeb, H., 2006. On the large-sample minimal coverage probability of confidence intervals after model selection. Journal of the American Statistical Association 101, 619--629.

\smallskip

\rf Maddala, G.S., 1971.  The use of variance components models in pooling cross section and time series data.  Econometrica 39, 341--358.

\smallskip

\rf Maddala, G.S., Mount, T.D., 1973.  A comparative study of alternative estimators for variance component models used in econometric applications.  Journal of the American Statistical Association 68, 324--328.

\smallskip

\rf Mann, V., De Stavola, B.L., Leon, D.A., 2004.  Separating within and between effects in family studies: an application to the study of blood pressure in children. Statistics in Medicine 23, 2745--2756.

\smallskip

\rf Rabe-Hesketh, S., Skrondal, A., 2012. Multilevel and Longitudinal Modeling Using Stata, 3rd edition. Stata Press, Texas.

\smallskip

\rf  Wooldridge, J. M., 2002.  Econometric Analysis of Cross Section and Panel Data.  MIT Press, Cambridge.

\newpage

\begin{figure}[p!]
 \centering
 \includegraphics[scale=0.7]{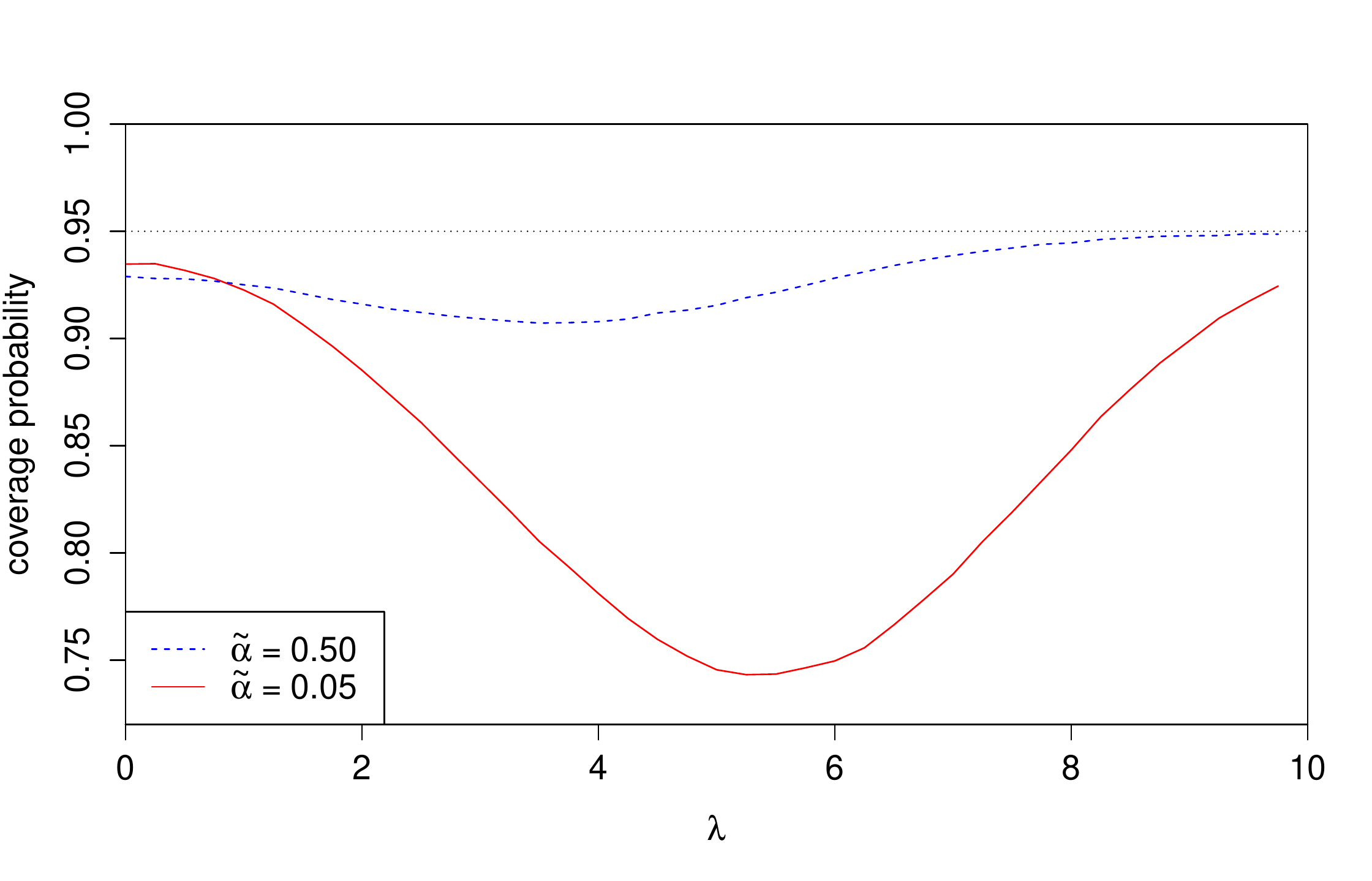}
 \caption{Graphs of the coverage probability functions of the confidence interval resulting from the two-stage procedure, when the
 usual unbiased estimators of the random error and random effect variances are used.
 Here $\lambda = N^{1/2}\tau$, where $\tau$ is the non-exogeneity parameter.
 The bottom and middle graphs are for nominal levels of significance, $\widetilde{\alpha}=0.05$ and $\widetilde{\alpha}=0.5$, respectively of the Hausman pretest. The matrix $G$ has off-diagonal
 elements $\rho$,
 (compound symmetry)
 where $\rho = 0.3$. The number of individuals $N = 100$, the number of time points $T = 3$,  $\psi = (\text{random effect standard deviation})/(\text{random error standard deviation}) = 1/3$ and the nominal coverage probability $1-\alpha = 0.95$.}
 \label{fig:Fig1}
\end{figure}

\newpage

\begin{figure}[p!]
 \centering
 \includegraphics[scale=0.7]{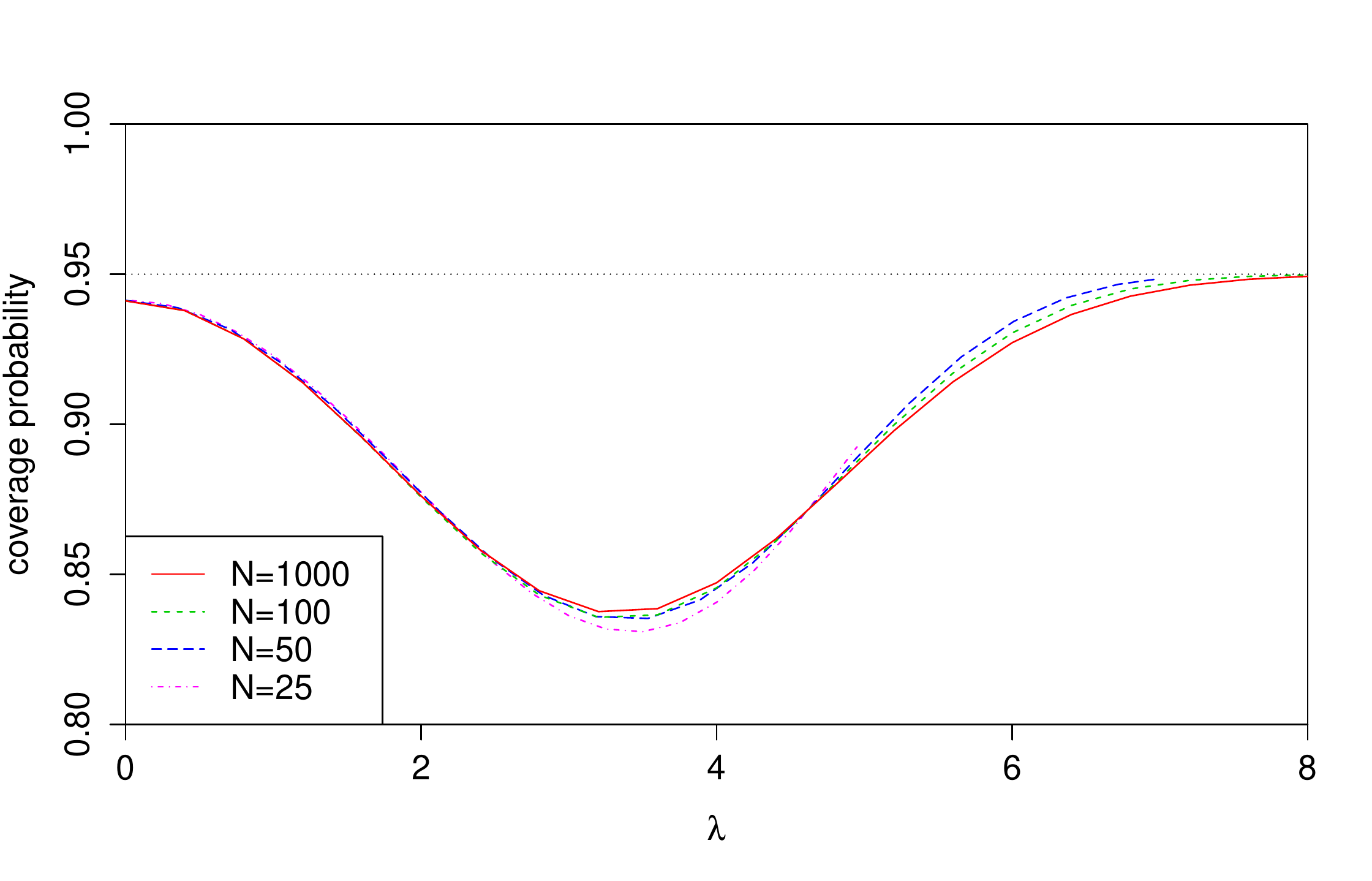}
 \caption{
 Graphs of the coverage probability functions of the confidence interval resulting from the two-stage procedure, when the
 usual unbiased estimators of the random error and random effect variances are used. Here, $\lambda = N^{1/2}\tau$, where $\tau$ is the non-exogeneity parameter,
 and $N = 25, 50, 100$ and 1000.
 The matrix $G$ has off-diagonal
 elements $\rho$, (compound symmetry) where $\rho = 0.4$. The number of time points  $T = 5$,  $\psi = (\text{random effect standard deviation})/(\text{random error standard deviation}) = 1/2$ and the nominal
nominal coverage probability $1-\alpha = 0.95$.
}
 \label{fig:Fig2}
\end{figure}
\clearpage

\begin{figure}[p!]
 \centering
 \includegraphics[scale=0.7]{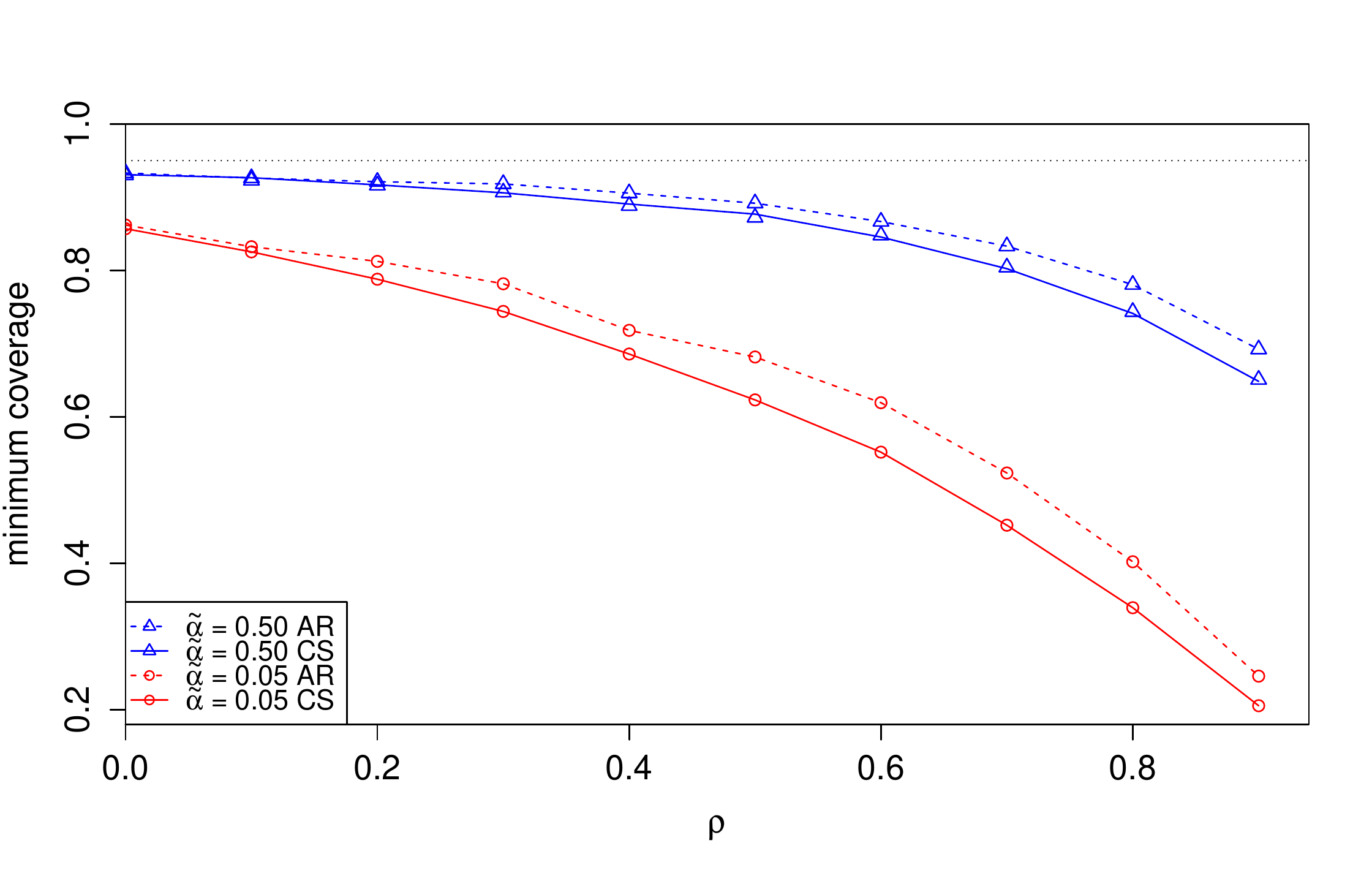}
 \caption{Graphs of the coverage probability functions, minimized over the non-exogeneity parameter $\tau$, of the confidence interval resulting from the two-stage procedure. This minimum coverage is considered as a function of $\rho$, for both compound symmetry (CS) and first order autoregression (AR) structures of the matrix $G$.
 The
 usual unbiased estimators of the random error and random effect variances are used.
 Two nominal levels of significance, $\widetilde{\alpha}=0.05$ and $\widetilde{\alpha}=0.5$, of the Hausman pretest are considered.
  The number of individuals $N = 100$, the number of time points $T = 3$,  $\psi = (\text{random effect standard deviation})/(\text{random error standard deviation}) = 1/3$ and the nominal coverage probability $1-\alpha = 0.95$.}
 \label{fig:Fig3}
\end{figure}
\clearpage

\begin{figure}[p!]
 \centering
 \includegraphics[scale=0.7]{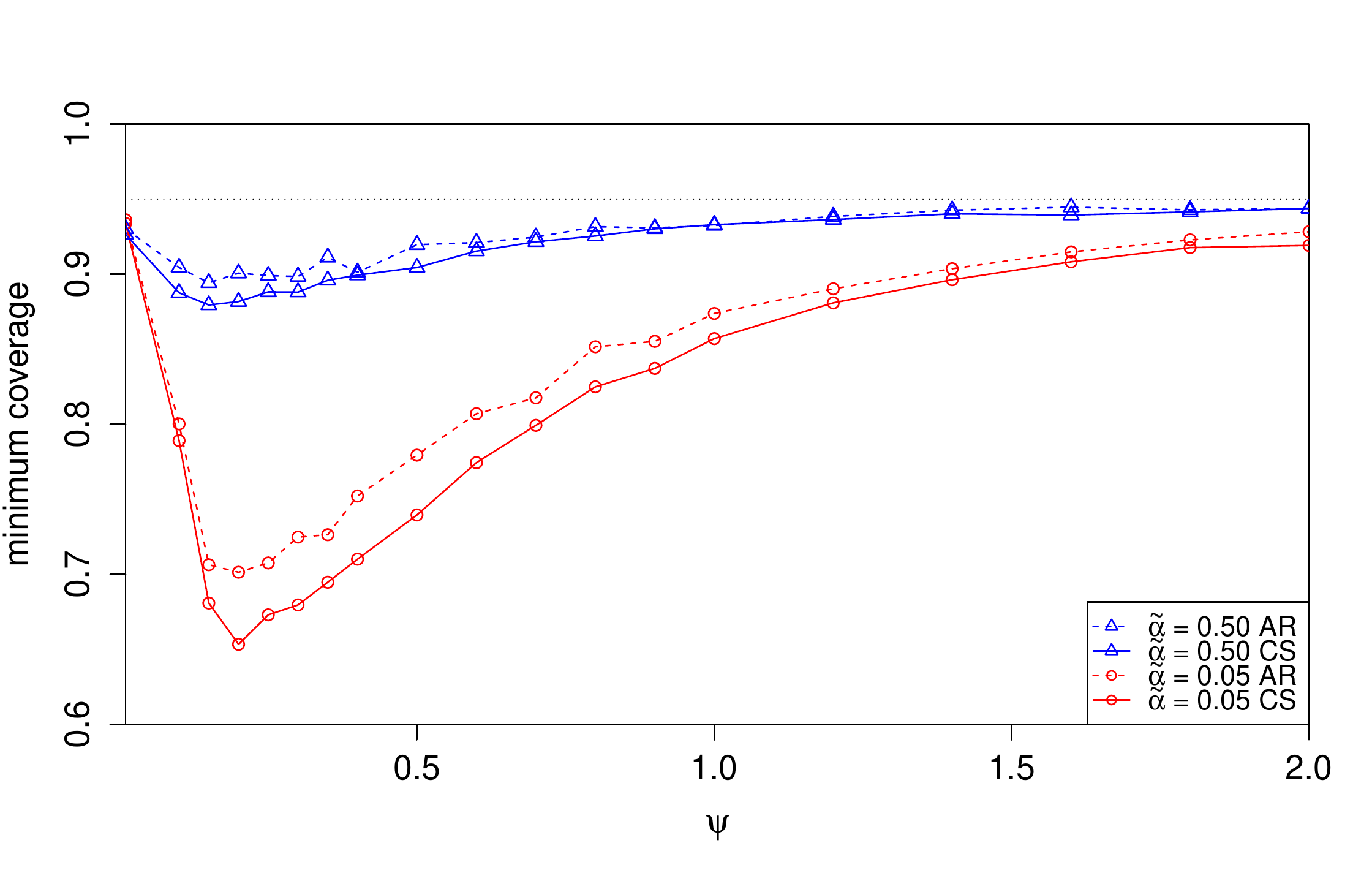}
 \caption{
Graphs of the coverage probability functions, minimized over the non-exogeneity parameter $\tau$, of the confidence interval resulting from the two-stage procedure. This minimum coverage is considered as a function of $\psi = (\text{random effect standard deviation})/(\text{random error standard deviation})$, for both compound symmetry (CS) and first order autoregression (AR) structures of the matrix $G$, where $\rho = 0.4$.
 The usual unbiased estimators of the random error and random effect variances are used.
 Two nominal levels of significance, $\widetilde{\alpha}=0.05$ and $\widetilde{\alpha}=0.5$, of the Hausman pretest are considered. The number of individuals $N = 100$, the number of time points $T = 3$ and the nominal coverage probability $1-\alpha = 0.95$.}
 \label{fig:Fig4}
\end{figure}

\end{document}